\documentclass[twocolumn]{aa}

\usepackage{amsmath,graphicx,txfonts,natbib,verbatim,amssymb}
\bibpunct{(}{)}{;}{a}{}{,}

\begin{document}
\titlerunning{Spatially resolved hard X-ray polarization in solar flares: effects of Compton scattering
and bremsstrahlung}
\authorrunning{Jeffrey and Kontar}

\title{Spatially resolved hard X-ray polarization in solar flares: effects of Compton scattering
and bremsstrahlung}

\author{N. L. S. Jeffrey \and E. P. Kontar}

\offprints{{N. L. S. Jeffrey \email{n.jeffrey@physics.gla.ac.uk}}}

\institute{School of Physics \& Astronomy, University of Glasgow, G12 8QQ, United Kingdom}

\date{Received ; Accepted}

\abstract{} {To study the polarization of hard X-ray (HXR) sources in the solar atmosphere,
including Compton backscattering of photons
in the photosphere (the albedo effect) and the spatial distribution of polarization across the
source. } {HXR photon polarization and spectra produced via electron-ion bremsstrahlung emission
are calculated from various electron distributions typical for solar flares. Compton scattering and
photoelectric absorption are then modelled using Monte Carlo simulations of photon transport in the
photosphere to study the observed (primary and albedo) sources. Polarization maps across HXR
sources (primary and albedo components) for each of the modelled electron distributions are
calculated at various source locations from the solar centre to the limb.} {We show that Compton
scattering produces a distinct polarization variation across the albedo patch at peak albedo
energies of 20-50 keV for all anisotropies modelled.  The results show that there are distinct
spatial polarization changes in both the radial and perpendicular to radial directions across the
extent of the HXR source at a given disk location. In the radial direction, the polarization
magnitude and direction at specific positions along the HXR source will either increase or decrease
with increased photon distribution directivity towards the photosphere. We also show how high
electron cutoff energies influence the direction of polarization at above $\sim$100 keV.}
{Spatially resolved HXR polarization measurements can provide important information about
the directivity and energetics of the electron distribution. Our results indicate the preferred
angular resolution of polarization measurements required to distinguish between the scattered and
primary components.  We also show how spatially resolved polarization measurements could be used to
probe the emission pattern of an HXR source, using both the magnitude and the direction of the
polarization.}


\maketitle

\section{Introduction}

Solar flare accelerated electrons can be studied via their radiation using various emission
mechanisms. Electrons with energies from tens to hundreds of keV can reach the denser layers of the
solar chromosphere and produce strong hard X-ray (HXR) emission via bremsstrahlung; the properties
of which tell us about the energetics and directivity of the accelerated electron distribution. A
major insight into the angular properties of the photon distribution and hence the radiating
electron distribution can come directly from the polarization of the HXRs. Since the anisotropy and
polarization of the photon distribution produced by bremsstrahlung will increase with the
anisotropy of the electron distribution \citep[e.g.][]{Brown1972, Haug1972, LeachPetrosian1983},
then in theory HXR polarization is one of the direct tools available to infer the anisotropy of the
emitting electron distribution. However, compared with the other HXR observables: energy, spatial
location, source size, time of arrival etc., polarization measurements through the years
\citep[see][as a review]{Kontaretal2011} have been fraught with difficulties and the measurements
obtained have often been met with skepticism. Nonetheless many missions have reported measurements
of HXR polarization from solar flares \citep{Tindoetal1970, Tindoetal1972, Nakadaetal1974,
Tindoetal1976, Lemenetal1982, McConnelletal2004, Boggsetal2006, SuarezGarciaetal2006, McConnelletal2007}.

Another valuable tool for determining the directivity of the photon distribution is the Compton
backscattered (albedo) component. Photons emitted towards the photosphere can Compton scatter back
out towards the observer and be detected alongside the primary photons directly emitted from the
source \citep{Tomblin1972, Santangeloetal1973}. Even for an isotropic source (minimum albedo), the
albedo component can account for as much as 40$\%$ of the observed flux at peak albedo energies of
20-50 keV \citep{BaiRamaty1978, Kontaretal2006} and for high downward beaming, albedo can
dominate the observed flux at certain energies. Separating the albedo component from the directly
emitted primary component is essential as it provides us with a unique (and only for individual
flares) opportunity of inferring the properties of the photon distribution directed towards the
photosphere \citep{KontarBrown2006,BattagliaKontar2011}, which otherwise would be poorly known. The
albedo component can also change various measurable properties of HXR sources, from producing a
distinctive change in the photon spectrum between $\sim10-100$ keV \citep{BaiRamaty1978} to
increasing source sizes and shifting the position of sources \citep{KontarJeffrey2010}. Compton
scattering is also polarization dependent and hence the polarization of an HXR source will be
altered by the albedo component \citep{Henoux1975,LangerPetrosian1977,BaiRamaty1978}. The albedo
component will not only change the polarization of the entire source but it will also produce
spatial variations in polarization across the extent of the albedo source.

Both HXR polarization and albedo measurements contribute information regarding the directivity of
the photon distribution and hence allow us to deduce the pitch angle distribution of the X-ray emitting
electrons \citep[e.g.][as a review]{Kontaretal2011}. In addition, the direction of the
polarization vector is related to the direction of electron beaming and hence the dominant
direction of electrons in a loop \citep{Emslieetal2008}. Spatially resolved polarization
measurements across an HXR source caused by albedo, have the advantage over spatially integrated
measurements because both the magnitude and the direction of polarization will change with photon
directivity and allow us to map the albedo and primary components. Therefore understanding how these
two parameters change with the photon anisotropy is essential and provides us with a new method of
investigating the entire photon anisotropy from a single HXR source.

Upcoming missions such as the Gamma-Ray Imager/Polarimeter for Solar flares (GRIPS)
\citep{Shihetal2009} (which is planned to be flown on a balloon in 2012) or other astrophysical
missions
\citep{Bloseretal2009,Mulerietal2009} may be able to provide us with
spatially resolved polarization measurements. Though GRIPS will have an angular resolution of $\sim
12''$ available at energies as low as $\sim25$ keV, it is doubtful that the GRIPS polarimeter will
be able to measure polarization at peak albedo energies of 20-50 keV, due to photoelectric
absorption dominating over Compton scattering in the detector at these energies, but hopefully some
technologies involved in GRIPS can be used for future missions. Nevertheless, although there are a
number of simulations for the spatially integrated polarization signal in flares
\citep{ElwertHaug1970,Haug1972,LangerPetrosian1977,BaiRamaty1978,Zharkovaetal1995,Emslieetal2008,Zharkovaetal2010},
there are currently no theoretical predictions or modelled spatially resolved hard X-ray
polarization maps. In addition, the effect of X-ray albedo on polarization signatures is often
ignored in the models.

In this paper, we compute spatially resolved polarization across HXR sources at various locations
on the solar disk, taking into account the influence of albedo, for various emitting electron
populations. We also examine how changing the maximum electron energy available during
bremsstrahlung can alter polarization measurements with photon energy, possibly providing a
diagnostic for the maximum electron cutoff energy.

\section{Bremsstrahlung interactions and the creation of the photon distribution}

\subsection{Stokes parameters and defining photon polarization}
The polarization state of incoherent radiation can be completely described using four Stokes
parameters \citep{Stokes1852,Chandrasekhar1960}. The Stokes vector consists of these four
parameters and takes the form of $S=[I, Q, U, V]$. The first Stokes parameter $I$ is the normalised
total intensity of the photon beam, while $Q/I$, $U/I$ and $V/I$ will have values between $-1$ and
$1$. The second and third normalised Stokes parameters are used to define linear polarization with $1$ or $-1$
indicating that the beam/photon packet is completely polarized with the sign providing the
direction of polarization. The fourth parameter is used to describe circular polarization but
bremsstrahlung emission in the solar corona, or Compton scattering only produces radiation that is
linearly polarized. In order to produce circularly polarised radiation via bremsstrahlung, the
spins of the radiating electrons need to be aligned and the magnetic field in the
corona/chromosphere is obviously not strong enough for this alignment. This means that only the
first three Stokes parameters are required and the fourth can be set to zero throughout our
simulations. Generally in X-ray and gamma ray astronomy the polarization of radiation is measured
using the degree of polarization (DOP) and the polarization angle $\Psi$, which is the preferred
direction of the electric field. These are defined using the Stokes parameters as,
\begin{equation}
DOP=\frac{\sqrt{Q^{2}+U^2}}{I},
\label{eq:DOP}
\end{equation}
and
\begin{equation}
\Psi=\frac{1}{2}\arctan\left(\frac{-U}{-Q}\right),
\label{eq:Psi}
\end{equation}
where the angle $\Psi$ is chosen to lie within the quadrant between $[-180^{\circ},180^{\circ}]$,
so that $\arctan\left(\frac{+0}{+0}\right)=+0, \arctan\left(\frac{+0}{-0}\right)=+180^{\circ}, \arctan\left(\frac{-0}{+0}\right)=-0$ and
$\arctan\left(\frac{-0}{-0}\right)=-180^{\circ}$. The negatives introduced into equation (\ref{eq:Psi}) ensure that a
negative $Q$ gives $0$ and a positive $Q$ gives $90^{\circ}$. Hence with this definition, when
$\Psi=0^{\circ}$, the observed radiation is polarized parallel to the radial direction at the solar
disk and when $\Psi=90^{\circ}$, the radiation is polarized perpendicular to the radial direction.
The opposite definition (i.e. $\Psi=90^{\circ}$ - polarized parallel to the radial direction) would
have been equally valid as long as definitions throughout the simulations are consistent.

We notice that the Stokes parameters are also frame dependent and hence have to be updated by the
use of rotation matrices \citep{Hovenier1983} when moving between different coordinate frames. The
rotations used in our simulations when moving between the source frame and the scattering frame and
vice versa are shown in appendices \ref{A.1.2} and \ref{app_B}. The DOP remains unchanged by the
rotation but the polarization angle $\Psi$ is measured with respect to the new frame.

For simplicity, the flare loop and hence the dominant direction of electrons is always assumed to
lie parallel with the local solar vertical, i.e. no tilt \citep[see][for non-zero
tilt]{Emslieetal2008}. This means that for the bremsstrahlung emission, the polarization direction
$\Psi$ across the entire source (source assumed to be small in comparison with the solar disk) and at
spatially resolved positions across the source can only ever equal $0^{\circ}$ or $90^{\circ}$,
 since $U$ from the bremsstrahlung emission is always close to zero \citep{BaiRamaty1978}.
Since in the solar disk frame (the observed frame of the HXR source), Compton scattering can
produce values of $U$ other than $0$, this means that the Compton scattered $\Psi$ can have values
other than $0^{\circ}$ or $90^{\circ}$. This is true for the spatially resolved polarization of the
albedo (and hence the total observed) source, while the spatially integrated albedo Stokes
parameters again sum to produce $\Psi$ values of either $0^{\circ}$ or $90^{\circ}$, for a
non-tilted loop.

\subsection{Primary photon distribution}

The intensity of the photon distribution $I(\epsilon,\theta)$ produced by bremsstrahlung for a
chosen electron distribution $F(E,\beta)$ is given by
\begin{eqnarray}
I(\epsilon,\theta)\propto\int^{\infty}_{E=\epsilon}\int^{2\pi}_{\Phi=0}\int^{\pi}_{\beta=0}F(E,\beta)
\sigma_{I}(E,\epsilon,\Theta) \sin\beta d\beta d\Phi dE
\label{eq:I}
\end{eqnarray}
where $\epsilon$ is the photon energy, $E$ is the electron energy, $F(E,\beta)$ is the target electron
flux density differential in energy
and $\sigma_{I}(E,\epsilon,\Theta)$ is the total (averaged over all polarization
states) angle-dependent bremsstrahlung cross-section
\citep{ElwertHaug1970,BaiRamaty1978,Massoneetal2004}. $\theta\in[0^{\circ},180^{\circ}]$ is the
photon polar emission angle measured from the local solar vertical with $\theta=0^{\circ}$ directed
away from the Sun. $\beta\in[0^{\circ},180^{\circ}]$ is the pitch angle of the emitting electrons
velocity to the local magnetic field with $\beta=0^{\circ}$ also directed away from the Sun.
$\Phi\in[0^{\circ},360^{\circ}]$ is the corresponding electron azimuthal angle and
$\Theta(\beta,\Phi,\theta)$ is the angle between the plane of emission (at angle $\beta$)
and the plane of observation (at angle $\theta$). The photon emission
is described by $\mu=\cos\theta$, where $\mu$ from 0 to 1 corresponds to emission away from the
Sun, and $\mu$ from -0 to -1 corresponds to emission towards the solar surface. The photon emission
angle $\mu=\cos\theta$ is related to the electron pitch angle $\beta$ by:
\begin{equation}
\cos\Theta=\cos\theta\cos\beta+\sin\theta\sin\beta\cos\Phi.
\end{equation}
Viewing the outward emission from $\mu=0.0-1.0$ corresponds to observing the source at a selected
heliocentric angle on the solar disk, i.e. $\mu=0.0$ would correspond to $90^{\circ}$ and is
equivalent to viewing a source sitting at the solar limb. The other Stokes parameters $Q$ and $U$
can be calculated in a similar manner:
\begin{equation}
Q(\epsilon,\theta)\propto\int^{\infty}_{E=\epsilon}\int^{2\pi}_{\Phi=0}\int^{\pi}_{\beta=0}F(E,\beta)\sigma_{Q}(E,\epsilon,\Theta)
 \sin\beta d\beta d\Phi dE,
\label{eq:Q}
\end{equation}
\begin{equation}
U(\epsilon,\theta)\propto\int^{\infty}_{E=\epsilon}\int^{2\pi}_{\Phi=0}\int^{\pi}_{\beta=0}F(E,\beta)\sigma_{U}(E,\epsilon,\Theta)
\sin\beta d\beta d\Phi dE ,
\label{eq:U}
\end{equation}
\\
with the only difference being the use of either $\sigma_{I}(E,\epsilon,\Theta)$,
$\sigma_{Q}(E,\epsilon,\Theta)$ or $\sigma_{U}(E,\epsilon,\Theta)$.
$Q(\epsilon,\theta)$ and $U(\epsilon,\theta)$ are then normalised between [-1,1] by dividing
through by $I(\epsilon,\theta)$. $\sigma_{I}(E,\epsilon,\Theta)$, $\sigma_{Q}(E,\epsilon,\Theta)$ and $\sigma_{U}(E,\epsilon,\Theta)$ are the polarization
dependent cross-sections for bremsstrahlung taken from \cite{GlucksternHull1953} and also following
the form used in \cite{Haug1972} and \cite{Emslieetal2008}, are given by
\begin{equation}
\sigma_{I}(E,\epsilon,\Theta)=\sigma_{\perp}(E,\epsilon,\Theta)+\sigma_{\parallel}(E,\epsilon,\Theta),
\end{equation}
\begin{equation}
\sigma_{Q}(E,\epsilon,\Theta)=(\sigma_{\perp}(E,\epsilon,\Theta)-\sigma_{\parallel}(E,\epsilon,\Theta))\cos2\Theta ,
\end{equation}
and
\begin{equation}
\sigma_{U}(E,\epsilon,\Theta)=(\sigma_{\perp}(E,\epsilon,\Theta)-\sigma_{\parallel}(E,\epsilon,\Theta))\sin2\Theta,
\end{equation}
where $\sigma_{\perp}(E,\epsilon,\Theta)$ and $\sigma_{\parallel}(E,\epsilon,\Theta)$ are the perpendicular and parallel components of the
bremsstrahlung cross-section respectively.

\subsubsection{Electron distribution}

In our simulations the X-ray emitting electron distribution is chosen to have the form,
\begin{equation}
F(E,\beta)\propto E^{-\delta}\exp\left({-\frac{(1+\cos\beta)^{2}}{\Delta\nu^{2}}}\right).
\label{eq:F_E_beta}
\end{equation}
The energy dependence follows a power law as shown by observations \citep[see][as recent
reviews]{Holmanetal2011,Kontaretal2011} and the electron angular distribution is a Gaussian. The
use of a Gaussian allows the angular anisotropy of the electron distribution to be easily
controlled by a single parameter $\Delta\nu$. The smaller the value of $\Delta\nu$, the greater the
proportion of the electron distribution, and hence the resulting photon emission directed
towards the photosphere.

\begin{figure*}
\centering
\includegraphics[width=14cm]{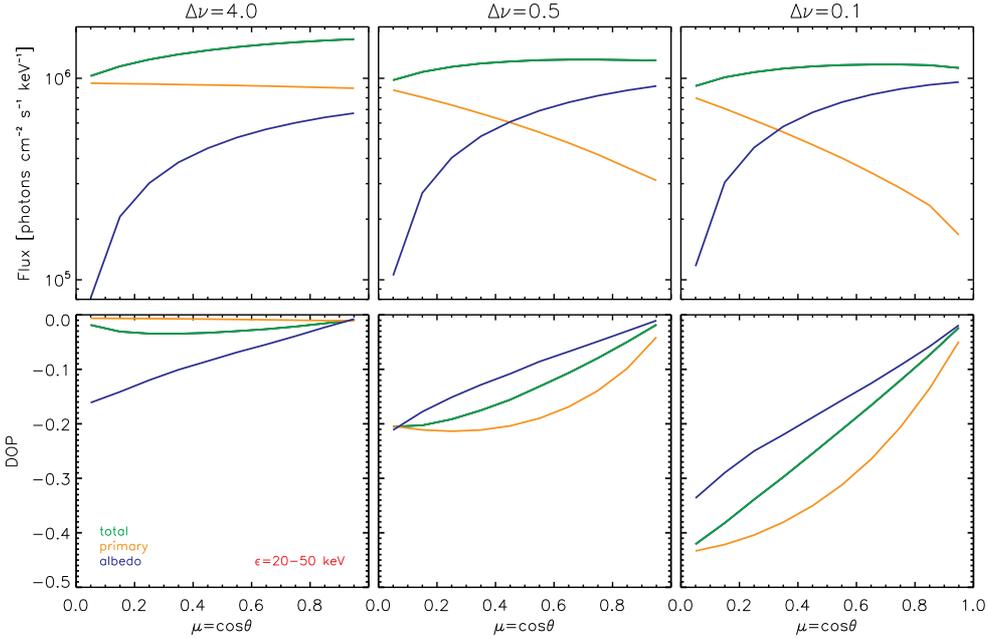}
\caption{Photon flux (top row) and DOP (bottom row) for upward primary (orange),
albedo (blue) and total (green) components across the entire source for both the $\Delta\nu= 4.0$
(1st column), the $\Delta\nu= 0.5$ (2nd column) and the $\Delta\nu=0.1$ (last column) created
photon distributions respectively, at the peak albedo energies of 20-50 keV, for different
locations on the solar disk from the centre ($\mu=1.0$) to the limb ($\mu=0.0$). Note here that a
negative DOP denotes that the direction of polarization $\Psi$ across the source is parallel to the
radial direction. The DOP for $\Delta\nu=4.0$ photon distribution (near isotropic
distribution) shows $\sim$ the same result as \cite{BaiRamaty1978}, as expected.}
\label{fig:Fig1}
\end{figure*}

In all simulations shown in this paper, we used a typical solar flare with $\delta=2$ (spectral index of
the mean electron spectrum and
$\Delta\nu=4.0$ , $\Delta\nu=0.5$ or $\Delta\nu=0.1$ to describe various pitch angle distributions
of electrons. Note that a target electron distribution of $\delta=2$ is produced by an injected electron
distribution of $\delta_{inj}=\delta+2=4$ \citep[see][as a review]{Holmanetal2011}.
A $\Delta\nu=4.0$ electron distribution produces an approximately
isotropic, unpolarised
photon distribution, while the $\Delta\nu=0.5$ and $\Delta\nu=0.1$ electron distributions produce photon distributions with
greater and greater beaming towards the photosphere. The resulting fluxes and polarizations of each
component [primary towards the observer only (orange), albedo (blue) and the total observed
emission (green)] plotted against emission/heliocentric angle $\mu=\cos\theta$ are shown for each
of the $\Delta\nu=4.0$, $\Delta\nu=0.5$ and $\Delta\nu=0.1$ photon distributions in Figure
\ref{fig:Fig1}. The resulting photon distributions in Figure \ref{fig:Fig1} are shown for the
energies of 20-50 keV, where the albedo emission peaks. As expected, even though the
$\Delta\nu=0.1$ distribution has a greater downward beaming (and hence a smaller observer directed
emission) than the $\Delta\nu=0.5$ distribution, within this low energy band the difference in
anisotropy between the two primary distributions cannot be clearly seen, however differences
between the two distributions can be clearly observed in the DOP with disk location over this
energy range (second row Figure \ref{fig:Fig1}).

A HXR source is modelled at a chosen height $h$ in the chromosphere. For the HXR energy range, the
source is about 1 Mm above the photosphere in accordance with recent X-ray observations
\citep{Kontaretal2008,Pratoetal2009,SaintHilaireetal2010,MrozekKowalczuk2010,Kontaretal2010,BattagliaKontar2011}.
The source spatial extent is modelled using a two-dimensional circular Gaussian distribution,
\begin{equation}
I(x,y)\propto\exp\left(-\frac{x^{2}}{2d^{2}}-\frac{y^{2}}{2d^{2}}\right),
\label{eq:I_xy}
\end{equation}
where $d$ is the size of the HXR source. Source sizes will be defined by the (Gaussian) Full Width
Half Maximum (FWHM), given by FWHM$\equiv\sqrt{8\ln(2)}d$.

\section{Monte Carlo simulations of photon transport}

A Monte Carlo code developed by \cite{KontarJeffrey2010} was modified to include polarization
(see appendices A and B, for details). The energy, angular and polarization properties of the input
photon distributions for the MC code are determined via the chosen input electron distributions
given by Equation \ref{eq:F_E_beta}, while the spatial properties are determined via Equation \ref{eq:I_xy}.
To input $I(\epsilon,\theta)$, $Q(\epsilon,\theta)$ and $U(\epsilon,\theta)$
(Equations \ref{eq:I}, \ref{eq:Q} and \ref{eq:U})
into our MC simulations, distributions of energy $\epsilon$, angle $\theta$ and polarization $Q$ and $U$
are numerically created from $I(\epsilon,\theta)$, $Q(\epsilon,\theta)$ and $U(\epsilon,\theta)$.
Each photon azimuthal angle $\phi$ is simply drawn from a uniform, random distribution between $0$ and $2\pi$.
To achieve the best statistics possible, we use $10^8$ photons in every MC code run.
Since the photon angular distribution is measured from the local solar vertical, photons with
$\theta < 90^{\circ}$ escape from the solar atmosphere as the primary emission and
photons with $\theta \ge 90^{\circ}$ will travel towards the
photosphere, creating the albedo emission. Photons are allowed to move freely until they reach the
photosphere, where they interact with free or bound electrons. In the simulations the photosphere is just defined by having a Hydrogen number density
of $n_{H}=1.16\times10^{17}$ cm$^{-3}$ \citep{Vernazzaetal1981}. Here the photons can no longer
travel without interacting in some way with the photospheric material. Each photon moves a
step-size before an interaction and this is calculated using $ss=-l\ln \zeta_{step}$, where $l$ is the
photon mean free path and $\zeta_{step}\in[0,1]$. The photon mean free path is calculated by
$l=1/n_{H}\sigma_{total}$, where $\sigma_{total}=\sigma_{c}+\sigma_{a}$, the addition of
the Compton scattering cross-section (using the form shown in \cite{BaiRamaty1978}) and
photoelectric absorption cross section. This photon mean free path, $l$, is of the order 100 km. A
photon interaction with the photospheric medium can either be by Compton scattering or by
photoelectric absorption. For photons with energies less than $\sim10$ keV, photoelectric
absorption is the more probable process while scattering dominates above $\sim10$ keV (Figure
\ref{fig:cross-sections}). When a photon is absorbed, it is removed from the simulations.  For each
photon, one of the two processes is chosen by calculating the ratio of $\sigma_{c}/\sigma_{total}$.
Another random number $\zeta_{pick}$ is then sampled from a uniform distribution between 0 and 1.
If the ratio is greater than $\zeta_{pick}$ then the photon is Compton scattered and if the ratio
is less than $\zeta_{pick}$, then the photon is absorbed.  The curvature of the Sun is included in
these simulations and a photon exits the photosphere when it satisfies the condition
$z>z_{\bigodot}=\sqrt{R_{\bigodot}^{2}-x^{2}-y^{2}}-R_{\bigodot}$, where $R_{\bigodot}$ is the
radius of the Sun (taken to be $6.96\times10^{10}$ cm $\sim960''$). The extent of the albedo patch
is limited by properly modelling the curvature of the Sun. The photons are allowed to scatter
multiple times until they exit the photosphere or are removed via
absorption. Most photons will leave the photosphere during the first scatter, with subsequent scatterings
producing less and less photons (this is shown in terms of Green's functions in \cite{Kontaretal2006}).
Photons that exit the photosphere with $\cos\theta > 0$ can then be collected into selected angular
or energy bins corresponding to HXR sources sitting at any chosen heliocentric
angle on/above the solar disk. In these simulations, source positions and sizes of each component
(primary and albedo) and the total source are given by the first and second moments of each
intensity distribution \citep{KontarJeffrey2010}, which are the mean $(\bar{x},\bar{y})$ and
variance (var$_{x}$, var$_{y})$ respectively. Though the primary distribution is initially
Gaussian, the albedo (and hence total observed) distribution will have a complex shape that is no
longer Gaussian. To quantify the sizes we use `Gaussian' Full Width Half
Maximum (FWHM) defined as 2.35 times the square root of the second-moment of the spatial 
distribution (not the actual FWHM of the complex distribution) \citep[see][for details]{KontarJeffrey2010}.
This allows a simple comparison with measurable parameters. All FWHM values in this paper 
are measured in this way.

\subsection{Compton scattering}
Above $\sim$10 keV, Compton scattering is the dominant process in the photosphere. The polarization
dependent differential Compton scattering cross-section is given by \cite{KleinNishina1929}, but
the form used in these simulations is from \cite{McMaster1961} and \cite{BaiRamaty1978}
\begin{equation}
\frac{d\sigma_{c}}{d\Omega}=\frac{1}{2}r_{0}^{2}\left(\frac{\epsilon}{\epsilon_{0}}\right)^{2}{\bigg\lgroup}\frac{\epsilon}{\epsilon_{0}}+
\frac{\epsilon_{0}}{\epsilon}-\sin^{2}\theta_{S}{\bigg\lgroup}1-Q\cos2\phi_{S}
 -U\sin2\phi_{S}{\bigg\rgroup}{\bigg\rgroup},
\end{equation}
where $r_{0}=2.82\times10^{-13}$cm is the classical electron radius, $\epsilon_{0}$ is the energy
of the incoming photon, $\epsilon$ is the energy of the outgoing photon, $\theta_{S}$ is the polar
scattering angle, $\phi_{S}$ is the azimuthal scattering angle and $Q$ and $U$ are the linear
Stokes parameters respectively \citep{McMaster1961, BaiRamaty1978}. The maximum change in DOP
occurs when $\theta_{S}=90^{\circ}$ and no change in DOP occurs for a backscattering at
$180^{\circ}$. The azimuthal scattering angle $\phi_{S}$ also has a non uniform dependency on the
incoming polarization state (see Appendix \ref{app_B}). In the simulations, the Klein-Nishina cross-section is
multiplied by $Z_{photo}=1.18$ to take account of elements higher than Hydrogen that are present
within the photosphere. $Z_{photo}$ indicates the average atomic number and the
number of electrons per Hydrogen atom in the photosphere, and is given by
\begin{equation}
Z_{photo}=\frac{\sum\limits _{Z}Z10^{A_{Z}}}{10^{A_{H}}}
\end{equation}
where $A_{Z}$ is the $\log_{10}$ abundance of an element with atomic number Z relative to
Hydrogen while $A_{H}=12$ is the $\log_{10}$ abundance of Hydrogen \citep{Asplundetal2009}. More details regarding the updating of photon
parameters ($\theta_{S},\epsilon,\phi_{S},Q,U$) after Compton scattering are given in the
appendices (A and B).

\subsection{Photoelectric absorption of X-ray photons}
\begin{figure}
\centering
\includegraphics[width=80mm]{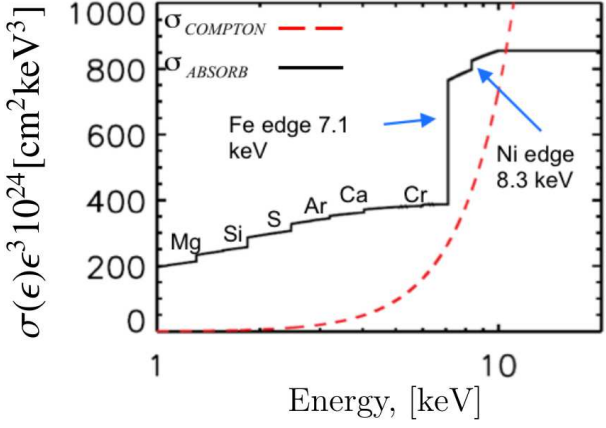}
\caption{Absorption  $\sigma_{a}$ (black solid) and Compton
$\sigma_{c}$ (red dashed) cross-sections
multiplied by $10^{24}\epsilon^{3}$ ($\epsilon$ in keV) at low energies below 10 keV. The absorption cross-section
is calculated using photospheric element abundances by \cite{Asplundetal2009}.}
\label{fig:cross-sections}
\end{figure}

For photons with energies below $\sim $10 keV, photoelectric absorption is the most probable photon
interaction in the photosphere. The process of absorption is heavily dependent upon the composition
of elements and the abundance of these elements within the photosphere. Absorption was therefore
modelled using the latest known solar photospheric abundances taken from \cite{Asplundetal2009}.
Absorption cross-section codes for the most important elements of H, He, C, N, O, Ne, Na, Mg, Al,
Si, S, Cl, Ar, Ca, Cr, Fe and Ni were adapted from \cite{Balucinska-ChurchMcCammon1992}. Absorption
is much less probable above $10$ keV and for energies higher than $10$ keV, the absorption
cross-section was approximated by $\sigma_a(\epsilon_{0})\propto\epsilon_{0}^{-3}$.  The
probability of photoelectric absorption
 is assumed to be independent of polarization \citep{Poutanenetal1996}.
Only the angular distribution of the ejected electron is dependent upon the photon polarization,
which is not modelled in our simulations. A comparison of the Compton scattering and absorption
cross sections is shown in Figure 2 (with $\sigma_{a}$ and $\sigma_{c}$ multiplied by $10^{24}
\epsilon^{3}$ ($\epsilon$ in keV) for comparison with \cite{MorrisonMcCammon1983}). Any differences between
Figure \ref{fig:cross-sections} and \cite{MorrisonMcCammon1983} are
due to the newer element abundances \citep{Asplundetal2009} and updated absorption cross-section codes
(in particular Helium) being used in our simulations.

\section{Spatial distribution of X-ray polarization}

\subsection{Single Compton scatter for an isotropic unpolarised source}

In order to demonstrate the spatial variation in polarization due to Compton scattering, the
easiest example to consider is the albedo patch created by an initially isotropic, unpolarised
point source at a height $h$ above the photosphere (see \cite{KontarJeffrey2010}). For this example, the variation in
polarization across the source can be described analytically by \citep{McMaster1961} and assuming
no energy losses,
\begin{equation}
{\rm DOP}=\frac{1-\cos^{2}\theta_{S}}{1+\cos^{2}\theta_{S}},
\label{eq:DOP_1scat}
\end{equation}
where
\begin{equation}
\cos\theta_{S}=\cos\theta\cos(\pi-\theta_{i})+\sin\theta\sin(\pi-\theta_{i})\cos\phi.
\label{eq:cosTheta_S}
\end{equation}
$\theta_{S}$ is the scattering angle, $\theta_{i}$ is the emission angle measured from the local
solar vertical, $\theta$ is the heliocentric angle on the solar disk and $\phi$ is the angle
measured in the solar disk plane. In equation (\ref{eq:DOP_1scat}), the scattering angle
$\theta_{S}$ determines the DOP. $\theta_{S}$ is related to distance $r$ from the centre of the
albedo patch by equation (\ref{eq:cosTheta_S}) though $r=h\tan(\pi-\theta_{i})$, and hence the DOP at any point across the albedo patch sitting at any heliocentric angle on the solar disk can be
easily calculated. Note that for this simple example, equation (\ref{eq:DOP_1scat}) assumes that
the energy difference between the incoming and scattered photon is negligible,
though this is only the case for low HXR energies of $\sim10$ keV. The DOP for all photon energies is calculated using the $T$ scattering matrix, which is used in the simulations (equation (\ref{eq:T_matrix}) in appendix \ref{app_B}) \citep{McMaster1961} .

\begin{figure}
\includegraphics[width=8cm]{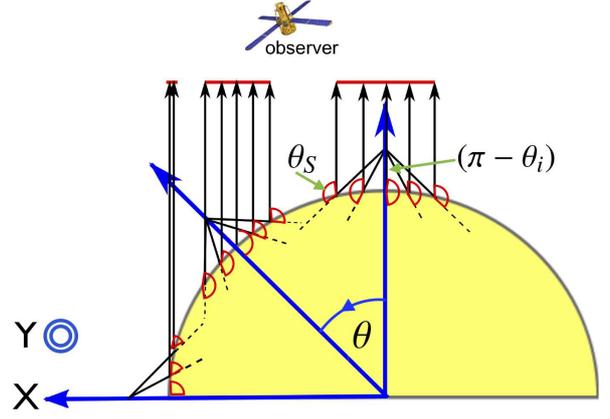}
\caption{Diagram of a single Compton scattering in the photosphere
for three heliocentric angles of $0^{\circ}$, $45^{\circ}$ and $90^{\circ}$. For a single scatter,
the DOP tends to $100\%$ as the scattering angle approaches $90^{\circ}$, producing a variation in
polarization across the extent of the source.}
\label{fig:cartoon1s}
\end{figure}

\begin{figure*}
\centering
\includegraphics[width=19cm]{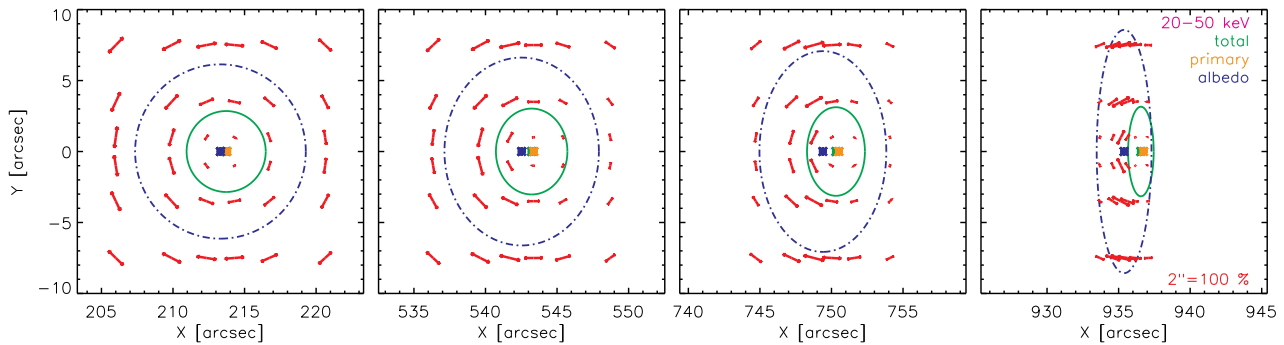}
\caption{Albedo polarization maps for an isotropic, unpolarised point source sitting above the
photosphere
at four different radial locations of $X=214'', 543'', 750'', 936''$ at $Y=0''$ (corresponding
$\mu=0.97,0.82,0.62,0.22$) after a single Compton scatter in the photosphere. All results are shown
at the peak albedo energies of 20-50 keV. The length of each red arrow indicates the DOP and the
direction of each arrow depicts the polarization angle $\Psi$ within the chosen plotting bin. The
solar radial direction (or $X$ axis for this case) is defined as the $\Psi = 0^{\circ}$ position. An
arrow length of $2''$ corresponds to a maximum DOP of $100 \%$. The green and blue ellipses give
the FWHM of the total and albedo sources respectively, while the green, blue and orange asterisks
give the centroid position of the total, albedo and primary sources.}
\label{fig:mapsss}
\end{figure*}

Figure \ref{fig:cartoon1s} shows a cartoon of this single scattering from an isotropic point source
for three different heliocentric angles of $0^{\circ}$, $45^{\circ}$ and $90^{\circ}$.  For a
source located exactly above the solar centre ($0^{\circ}$) at a height $h$, the resulting variation across the
photospheric albedo patch is radially symmetric. As the radial distance $r$ from the source centre
increases, the scattering angle $\theta_{S}$ of any observed radiation will decrease from $180^{\circ}$ towards $90^{\circ}$,
causing the DOP to grow from $0\%$ to $\sim 100\%$.
Radiation scattered in the photosphere at a location directly below the source ($180^{\circ}$ backscatter) and emitted
towards the observer will experience no change in its DOP. This
statement is true for sources at any heliocentric angle $\theta_{i}$ but
projection effects at angles $\theta_{i}>0^{\circ}$ will create an asymmetry in the polarization pattern along the radial direction,
whereas the polarization pattern in the perpendicular to radial direction always remains
symmetrical. The described pattern can be seen in Figure \ref{fig:mapsss} which shows the
polarization maps for a single Compton scattering at four disk locations
ranging from the solar centre to the limb, along $Y=0''$ at $X=214'' , 543'' , 750'', 936''$.
These locations are equivalent to $\mu=0.97,0.82,0.62,0.22$ ($\theta_{i}=14^{\circ},35^{\circ},52^{\circ},77^{\circ}$) and denote the approximate positions of the total source,
which are shifted from the primary position due to the albedo component \citep{KontarJeffrey2010}.
The polarization across the source at any location on the solar disk can always be measured with respect to the radial connecting the solar
disk centre and the centre of the source. Therefore due to the symmetry of the problem, source
locations at $Y=0''$ considered in this paper can straightforwardly be applied to any solar disk
location.  In Figure \ref{fig:mapsss}, the dotted blue ellipse denotes the FWHM of the albedo
component, the solid green ellipse denotes the FWHM of the total observed source and the orange,
blue and green asterisks indicate the centroid positions of primary, albedo and total observed
sources respectively. The polarization angle $\Psi$ follows a steady pattern across all
sources. $\Psi$ is always at an angle tangential to the line connecting the desired position and
the location of the source centroid position. Hence at disk centre locations the $\Psi$ pattern is also symmetrical
across the source.

\begin{figure*}
\centering
\includegraphics[width=19cm]{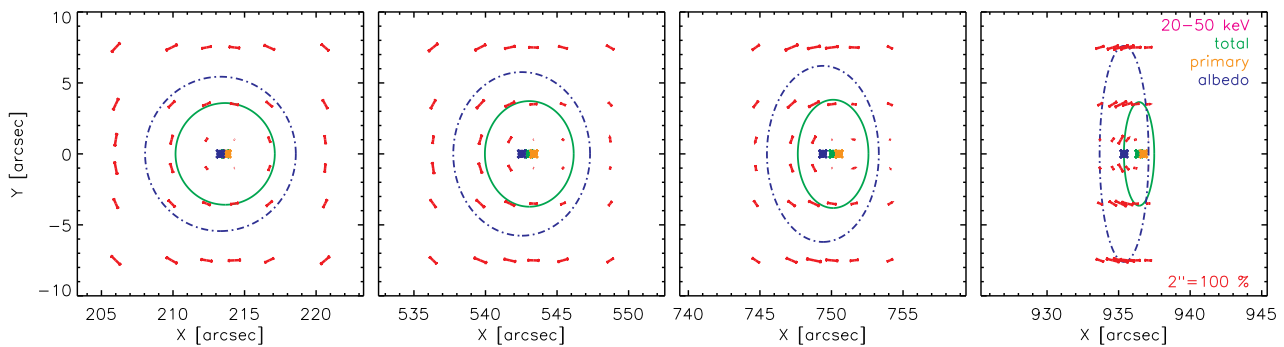}
\caption{Albedo polarization maps as in Figure \ref{fig:mapsss}, but for the case of multiple
Compton scatterings
in the photosphere. Multiple scatterings have} acted to decrease the DOP at all points across each
source.
\label{fig:mapsms}
\end{figure*}

Figure \ref{fig:mapsms} shows the same albedo
polarization maps as in Figure \ref{fig:mapsss} but for multiple Compton scatterings.
The overall pattern for the DOP and $\Psi$ are preserved but due to multiple scatterings the overall DOP across
all locations over the albedo patch has decreased. A single scattering DOP of $\sim100\%$ near the
edge of the source has been reduced to $\sim50\%$ by multiple scatterings. All other simulations
shown below are for multiple Compton scatterings.

The albedo pattern from a primary source at a greater
height than 1 Mm (say from a coronal source) should produce the same albedo polarization
pattern but over a much greater area in the photosphere. Plotting the polarization pattern at a higher or lower energy than
the peak albedo range of 20-50 keV would obscure the albedo polarization pattern due to lowering
albedo fluxes at these energies.

\subsection{Anisotropic source at a height of $h=1$ Mm ($1.4''$) and size of $5''$}

For a chosen chromospheric HXR source height of $h =1$ Mm $(1.4'')$ and a primary source size of
FWHM$\sim5''$, simulations were performed for all three photon distributions created by the
$\Delta\nu= 4.0$, $\Delta\nu= 0.5$ and $\Delta\nu= 0.1$ electron distributions. This height was
chosen to match chromospheric HXR source height measurements
\citep{Kontaretal2008,Pratoetal2009,SaintHilaireetal2010,MrozekKowalczuk2010,Kontaretal2010,BattagliaKontar2011}.
All the results shown here are for the energy range of 20-50 keV, where albedo emission peaks,
producing the largest distortion to the primary component but the best opportunity for the
detection of the albedo component.

\begin{figure*}
\includegraphics[width=19cm]{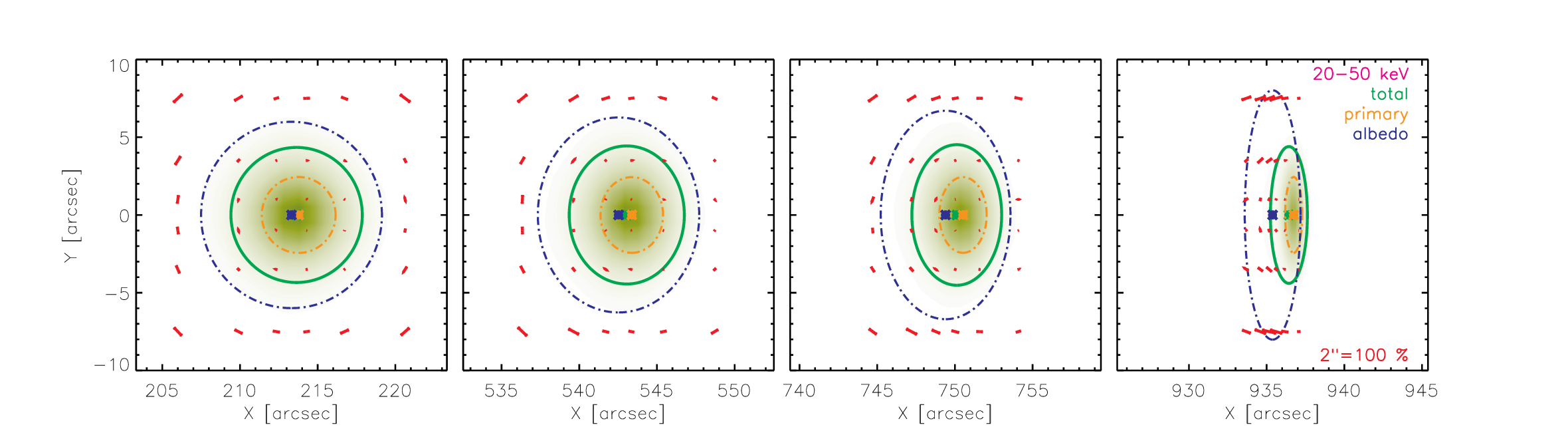}
\caption{Total X-ray brightness and polarization (arrows) maps for the photon distribution created
by the near isotropic $\Delta\nu=4.0$ electron distribution for a $5''$ primary source. The resulting total sources sit at four disk locations of $X=213.6'', 542.9'', 750.1'', 936.6''$ at $Y=0''$(corresponding to $\mu\sim0.97,0.82,0.62,0.22$). The green, blue and orange ellipses
and asterisks give the FWHM and centroid positions of the total, albedo and primary sources
respectively. The primary component is dominant at all disk locations and hence the albedo
polarization pattern at small radii (over the extent of the $5''$ primary source) is obscured
slightly by the flat, constant polarization of the primary source.}
\label{fig:maps_dnu4}
\end{figure*}

\begin{figure*}
\includegraphics[width=19cm]{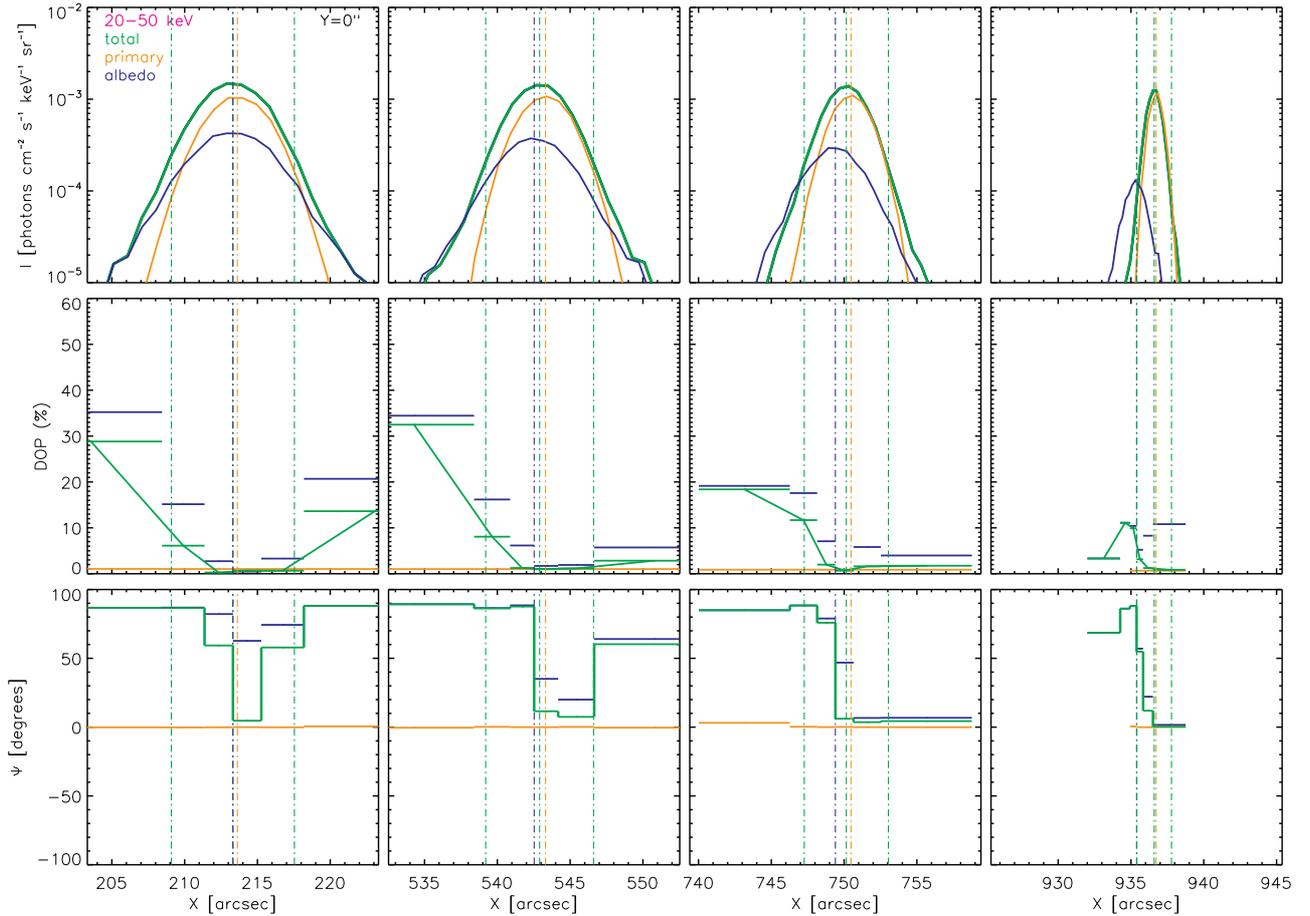}
\caption{Radial slices through each of the sources in Figure \ref{fig:maps_dnu4}
for the $\Delta\nu=4.0$ created photon distribution.
Each of the radial slices are taken along $X$ at $Y=0''$ for the intensity, $I$, DOP and
polarization angle $\Psi$ at the four source locations. As before, orange=primary,
blue=albedo and green=total. The green dash-dot lines denote the centroid position and the FWHM of
the total observed source while the orange and blue lines denote the centroid positions of the
primary and albedo sources respectively. Slices along the radial direction for the
$\Delta\nu=4.0$ distribution show a clear DOP and $\Psi$ pattern at all four disk locations
plotted.}
\label{fig:slicesX_dnu4}
\end{figure*}

\begin{figure*}
\includegraphics[width=19cm]{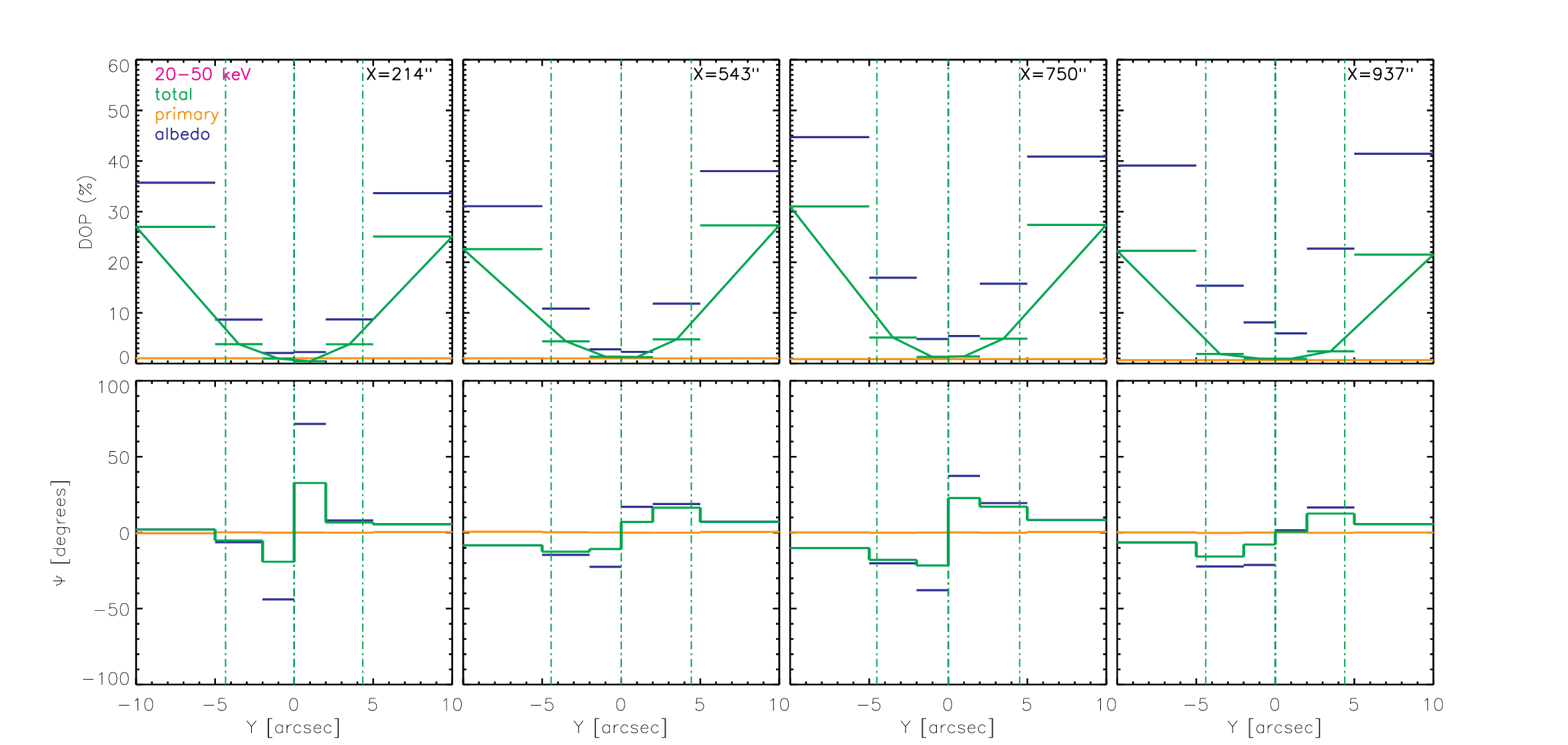}
\caption{Perpendicular to radial slices through each of the sources shown in Figure
\ref{fig:maps_dnu4} for the $\Delta\nu=4.0$ distribution.
Each of the perpendicular to radial slices are taken along $Y$  at $X=213.6'', 542.9'', 750.1'', 936.6''$ for the DOP and polarization angle $\Psi$. The lines and
colours are as in Figures \ref{fig:maps_dnu4}, \ref{fig:slicesX_dnu4}. The DOP and the magnitude
of the polarization angle $\Psi$ remain symmetrical along the source perpendicular to radial direction.}
\label{fig:slicesY_dnu4}
\end{figure*}

Figures \ref{fig:maps_dnu4}, \ref{fig:maps_dnu05} and \ref{fig:maps_dnu01} each plot the resulting polarisation maps for four HXR sources
(resulting from the primary and albedo components) created by the
$\Delta\nu= 4.0$, $\Delta\nu=0.5$ and $\Delta\nu=0.1$ electron distributions respectively. As with Figures \ref{fig:mapsss} and \ref{fig:mapsms},
each figure plots four HXR sources positioned at $\mu\sim0.97,0.82,0.62,0.22$.
In Figures \ref{fig:maps_dnu4}, \ref{fig:maps_dnu05} and \ref{fig:maps_dnu01}, the dotted ellipses denote the FWHM of the total source (green),
the primary source (orange) and the albedo source (blue) and
the correspondingly coloured asterisks denote the (x,y) centroid position of the total source and
the primary and albedo components respectively. Figures \ref{fig:slicesX_dnu4}, \ref{fig:slicesX_dnu05} and \ref{fig:slicesX_dnu01} plot intensity,
I (top row), DOP (middle row) and $\Psi$ (bottom row) along the radial
direction $X$ centred at Y $= 0''$ (across a of bin width$= 2''$) for each of the maps in Figures \ref{fig:maps_dnu4}, \ref{fig:maps_dnu05} and \ref{fig:maps_dnu01} . Figures
\ref{fig:slicesY_dnu4},  \ref{fig:slicesY_dnu05} and \ref{fig:slicesY_dnu01} plot the DOP (top row) and $\Psi$ (bottom row) along the perpendicular to radial direction $Y$
centred at X$=213.6'' ,543.9'' ,750.1'' ,936.6''$ (again across a bin width$=2''$ ) for each of the maps in Figures \ref{fig:maps_dnu4}, \ref{fig:maps_dnu05} and \ref{fig:maps_dnu01}.


\subsection{Quasi-isotropic distribution with $\Delta\nu=4.0$}

The photon distribution produced by the $\Delta\nu=4.0$ electron distribution is approximately
unpolarised and isotropic. Therefore, both spatially integrated and spatially resolved polarization measurements across the primary source at all locations on the solar disk produce
DOP$\sim 0\%$ and $\Psi=0^{\circ}$ (radial) at $20-50$ keV.
The albedo component produces asymmetrical DOP and $\Psi$ variations along the source radial direction (Figure \ref{fig:slicesX_dnu4})
while along the source perpendicular to radial direction (Figure \ref{fig:slicesY_dnu4}),
variations in albedo DOP and $\Psi$ are approximately symmetrical, since the centroid positions of the primary and albedo components always coincide in the perpendicular to radial direction.
The simulated HXR sources plotted in Figure \ref{fig:maps_dnu4} have a finite source size of $\sim5''$.
Compared with a point source (Figure \ref{fig:mapsms}), this produces two main differences:
i) photons leave the source from different positions above the photosphere and
ii) for certain distributions and disk locations, the primary polarization will dominate over the extent of the primary source.
The first property reduces the DOP at all points across the source, compared with the albedo patch created by a point source. Due to the second property, the polarization variation caused by albedo may be slightly masked by the primary component within the source FWHM, especially for cases where the primary component is dominant, i.e. isotropic or near isotropic distributions.

For the quasi-isotropic $\Delta\nu=4.0$ distribution, the primary component is the dominant component at
all four disk locations, with the albedo contribution falling as the source location moves towards
the limb (Figure \ref{fig:slicesX_dnu4}- first row). Hence, the primary component dominates within the
FWHM of the total source, while the albedo component dominates after this boundary.
The second and third rows of Figure \ref{fig:slicesX_dnu4} demonstrate common radial trends in DOP
and $\Psi$, not only across the extent of each individual source but also between sources at different disk locations.
For a quasi-isotropic distribution, at a particular disk location (other than the disk centre), the highest DOP along the radial direction is observed at the disk-centre-side of the source (where the albedo dominates). This falls to approximately zero within the FWHM of the total source (where the primary dominates) and then increases again towards the limb-side of the source (where the albedo again dominates), but always remains lower than the DOP at the disk-centre-side. Comparing the four disk locations, the DOP
at all points along the radial direction of a single source decreases as the source location nears the limb.
In the radial direction, spatially resolved DOP can achieve values as high as $\sim30\%$ at disk centre locations.
The albedo component produces the distinctive $\Psi$ variation shown in Figure \ref{fig:maps_dnu4}.
Along the radial direction, $\Psi=90^{\circ}$ at the disk-centre-side of the source, falls to zero within the FWHM extent and then rises again at the limb-side of the source.
Along the perpendicular to radial direction of a single source (Figure \ref{fig:slicesY_dnu4}), DOP (first row) and the magnitude of $\Psi$ (second row) are symmetrical due to no projection effects and the centres of the primary and albedo components always coinciding. As with the radial direction, in the perpendicular to radial direction, spatially resolved DOP can achieve values as high as $\sim30\%$.
While the magnitude of $\Psi$ across the
source is symmetrical at each disk location in the perpendicular to radial direction,
$\Psi$ itself behaves as an odd function along Y, with a $180^{\circ}$ rotational symmetry about the source centre, increasing from the radial at the upper source edge to $\left|{\Psi>0^{\circ}}\right|$ and then back to the radial direction at the lower source edge.

\subsection{Beamed electron distributions $\Delta\nu=0.5$ and $\Delta\nu=0.1$}

\begin{figure*}
\includegraphics[width=17cm]{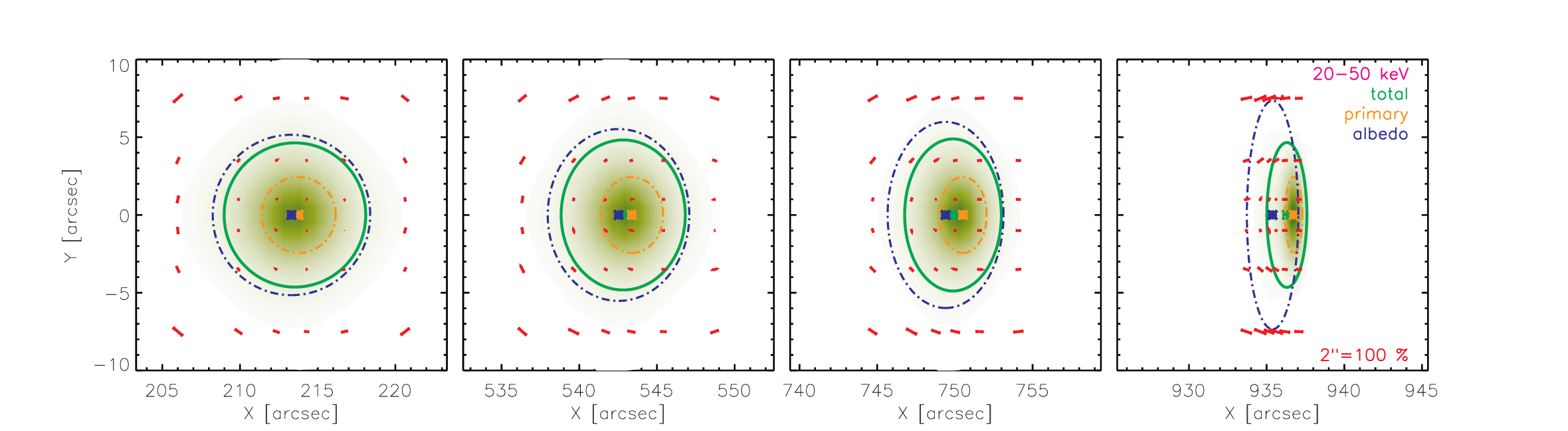}
\caption{Total X-ray brightness and polarization maps for the photon distribution created by the $\Delta\nu=0.5$
electron distribution for a $5''$ primary
source. The resulting total sources sit at four disk locations of $X=213.1'', 542.6'', 750.0'', 936.5''$ at $Y=0''$ (corresponding to $\mu\sim0.97,0.82,0.62,0.22$). The green, blue and orange ellipses give the FWHM of the
total, albedo and primary sources respectively.  Green, blue and orange asterisks give the centroid
position of the total, albedo and primary sources. For disk locations closer to the solar centre,
the albedo component is the dominant component over the primary component.}
\label{fig:maps_dnu05}
\end{figure*}

\begin{figure*}
\includegraphics[width=17cm]{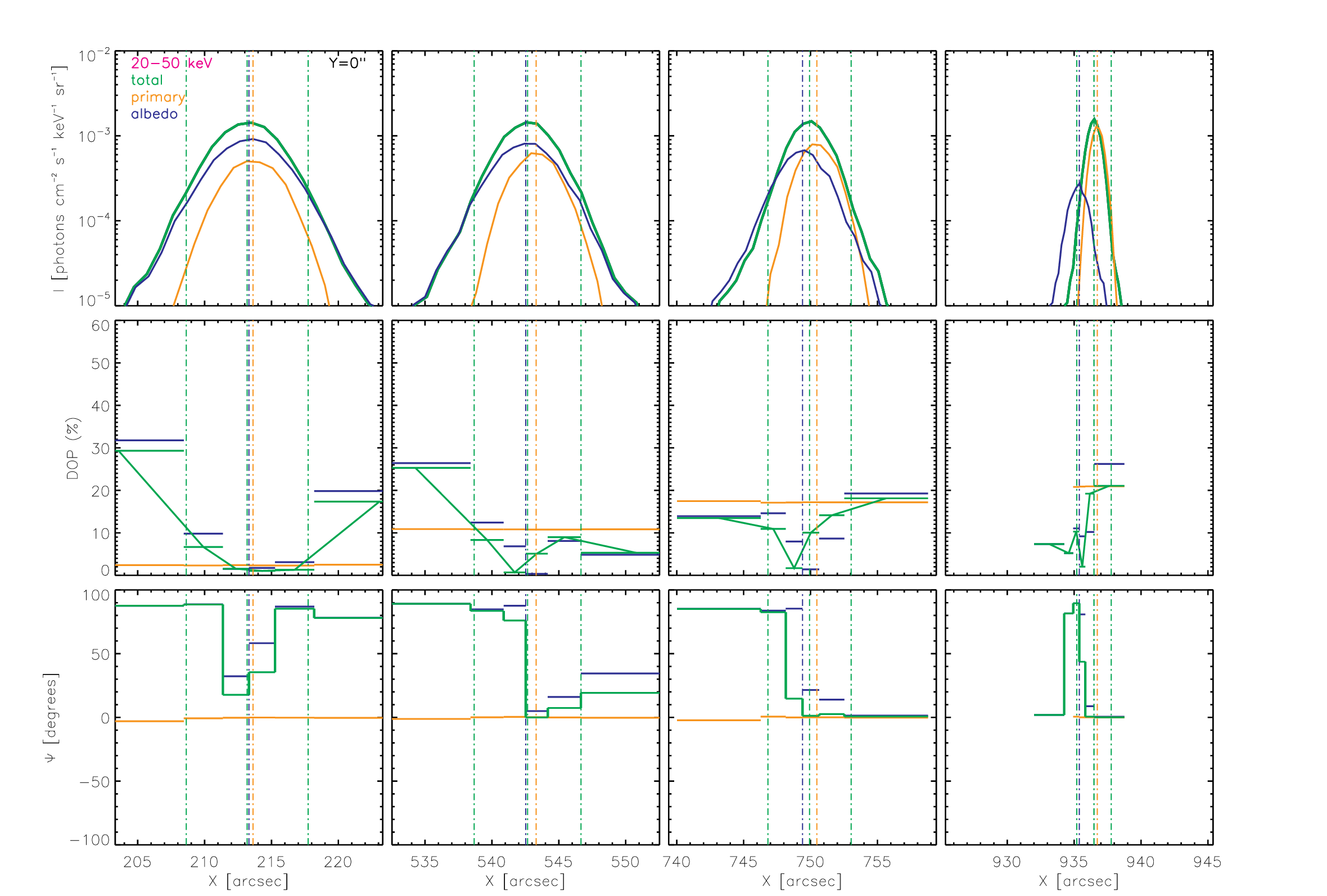}
\caption{Radial slices (along $X$) through $Y=0''$ for the intensity, $I$, the DOP and $\Psi$
for each of the sources in Figure \ref{fig:maps_dnu05} for the $\Delta\nu=0.5$ distribution. Again, orange=primary, blue=albedo and green=total. The green dash-dot
lines denote the centroid position and the FWHM of the total observed source while the orange and
blue lines denote the centroid positions of the primary and albedo sources respectively.}
\label{fig:slicesX_dnu05}
\end{figure*}

\begin{figure*}
\includegraphics[width=17cm]{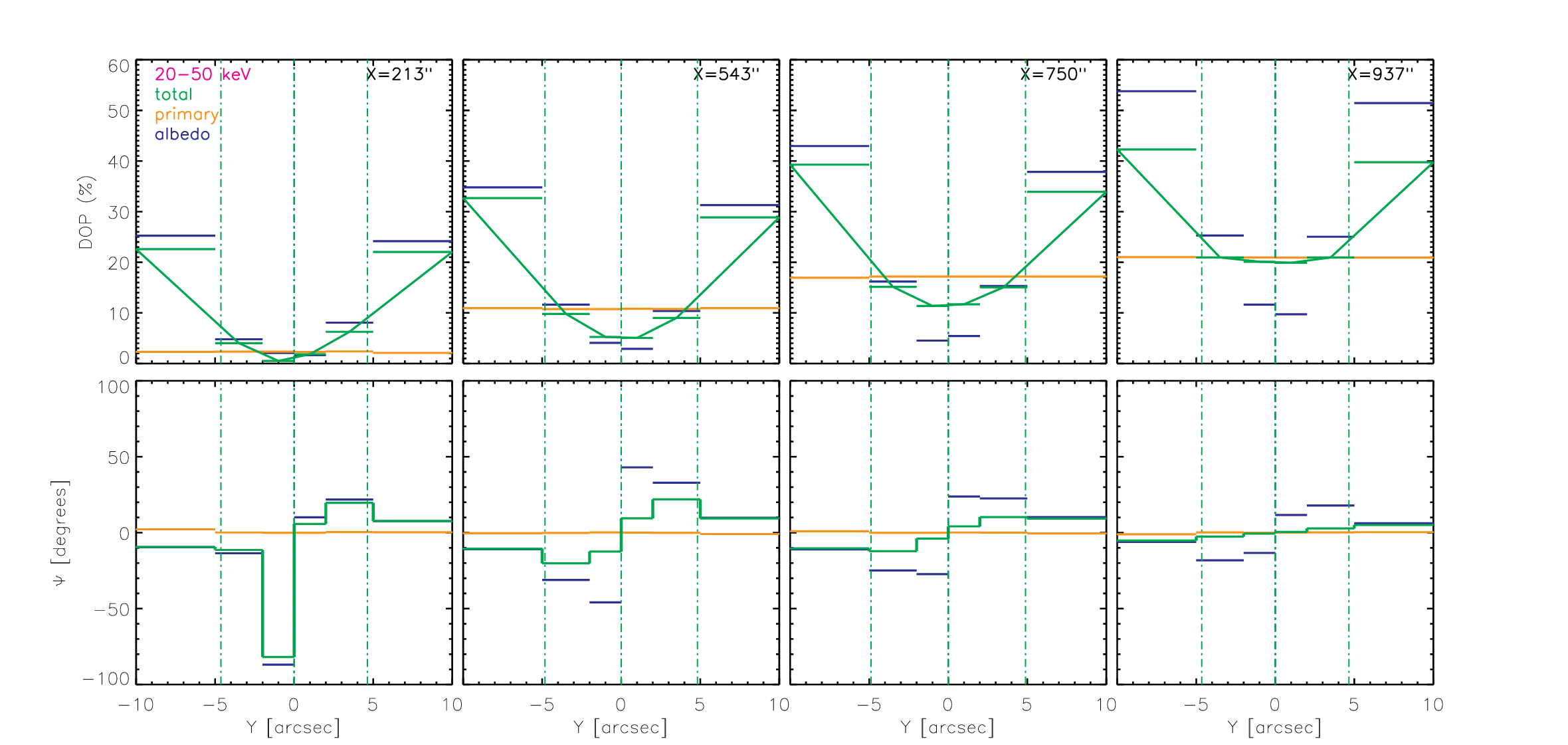}
\caption{Perpendicular to radial slices through each of the sources shown in Figure
\ref{fig:maps_dnu05} for the $\Delta\nu=0.5$ distribution.
Each of the perpendicular to radial slices are taken along $Y$  at $X=213.1'', 542.6'', 750.0'', 936.5''$ for the DOP and polarization angle $\Psi$. The lines and
colours are as in Figures \ref{fig:maps_dnu05}, \ref{fig:slicesX_dnu05}.}
\label{fig:slicesY_dnu05}
\end{figure*}

\begin{figure*}
\includegraphics[width=17cm]{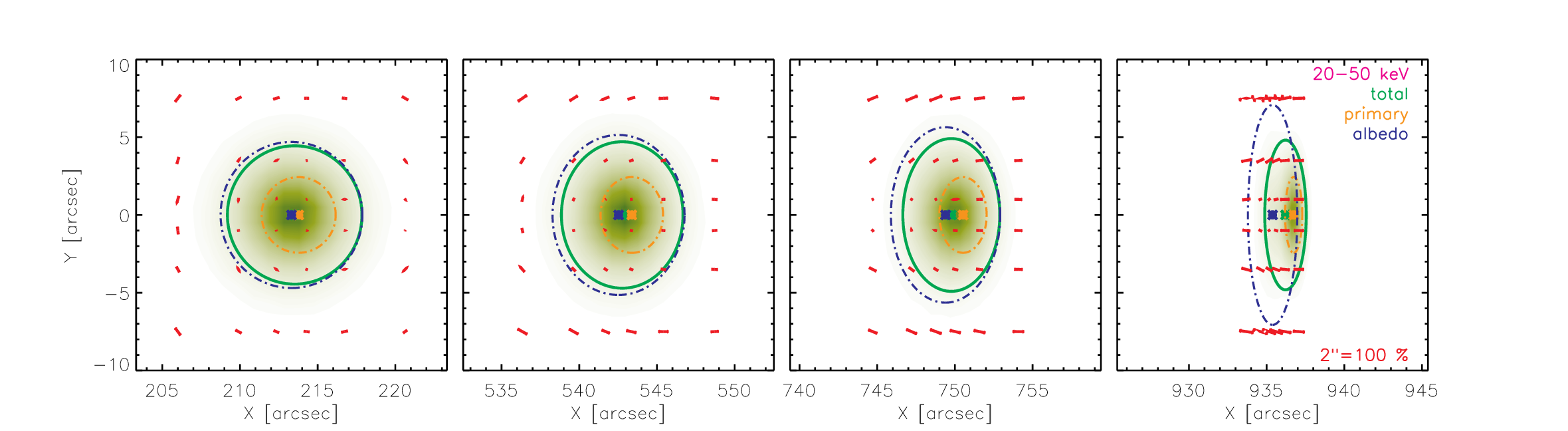}
\caption{Total X-ray brightness and polarization maps for the photon distribution created by the $\Delta\nu=0.1$
electron distribution for a $5''$ primary
source. The resulting total sources sit at four disk locations of $X=213.0'', 542.5'', 749.9'', 936.4''$ at $Y=0''$ (corresponding to $\mu\sim0.97,0.82,0.62,0.22$). The green, blue and orange ellipses give the FWHM of the
total, albedo and primary sources respectively.  Green, blue and orange asterisks give the centroid
position of the total, albedo and primary sources. At all disk locations the albedo
component is the dominant component over the primary component.}
\label{fig:maps_dnu01}
\end{figure*}

\begin{figure*}
\includegraphics[width=17cm]{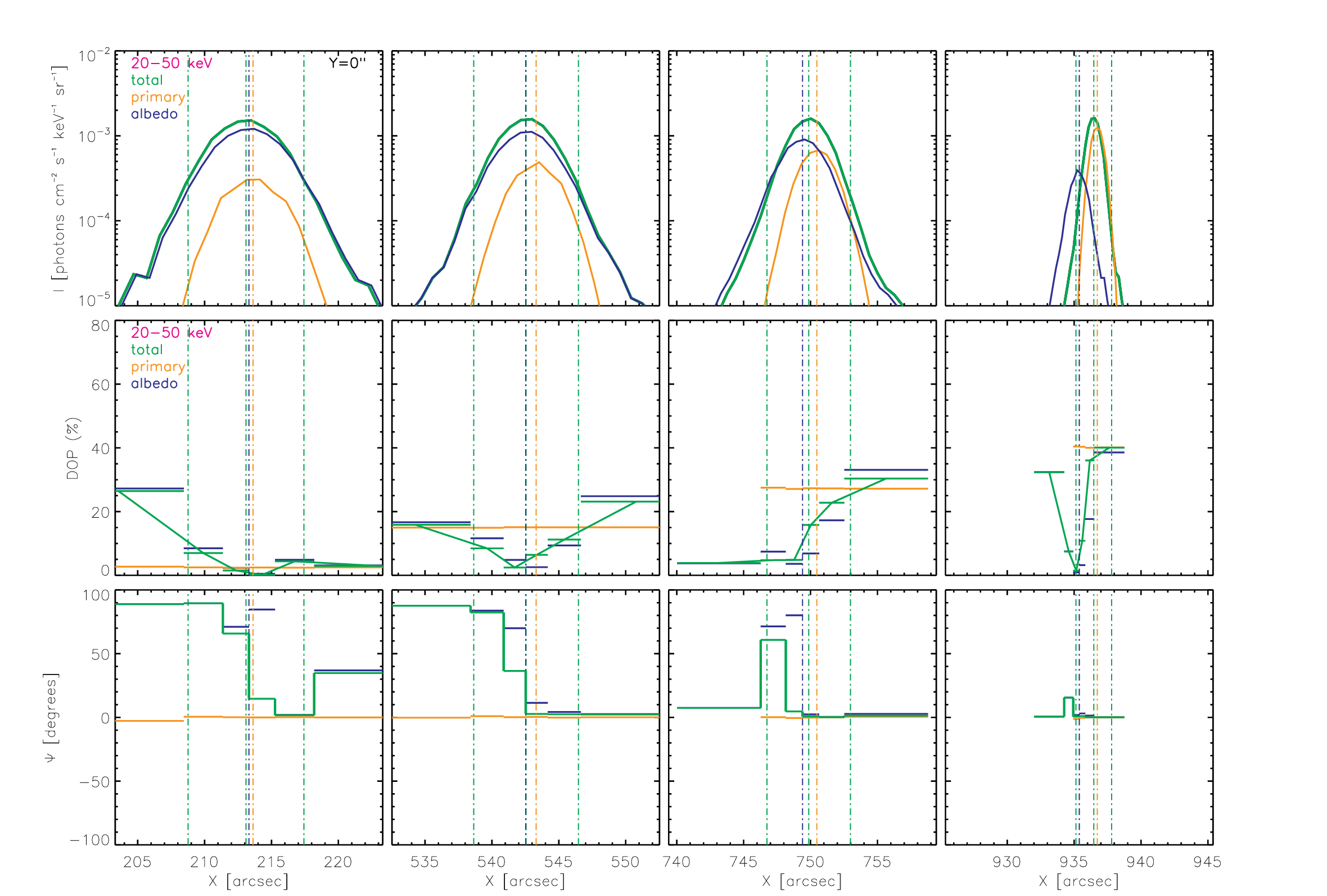}\caption{Radial slices (along $X$) through $Y=0''$ for the intensity, $I$, the DOP and $\Psi$
for each of the sources in Figure \ref{fig:maps_dnu01} for the $\Delta\nu=0.1$ distribution. Again, orange=primary, blue=albedo and green=total. The green dash-dot
lines denote the centroid position and the FWHM of the total observed source while the orange and
blue lines denote the centroid positions of the primary and albedo sources respectively.}
\label{fig:slicesX_dnu01}
\end{figure*}

\begin{figure*}
\includegraphics[width=17cm]{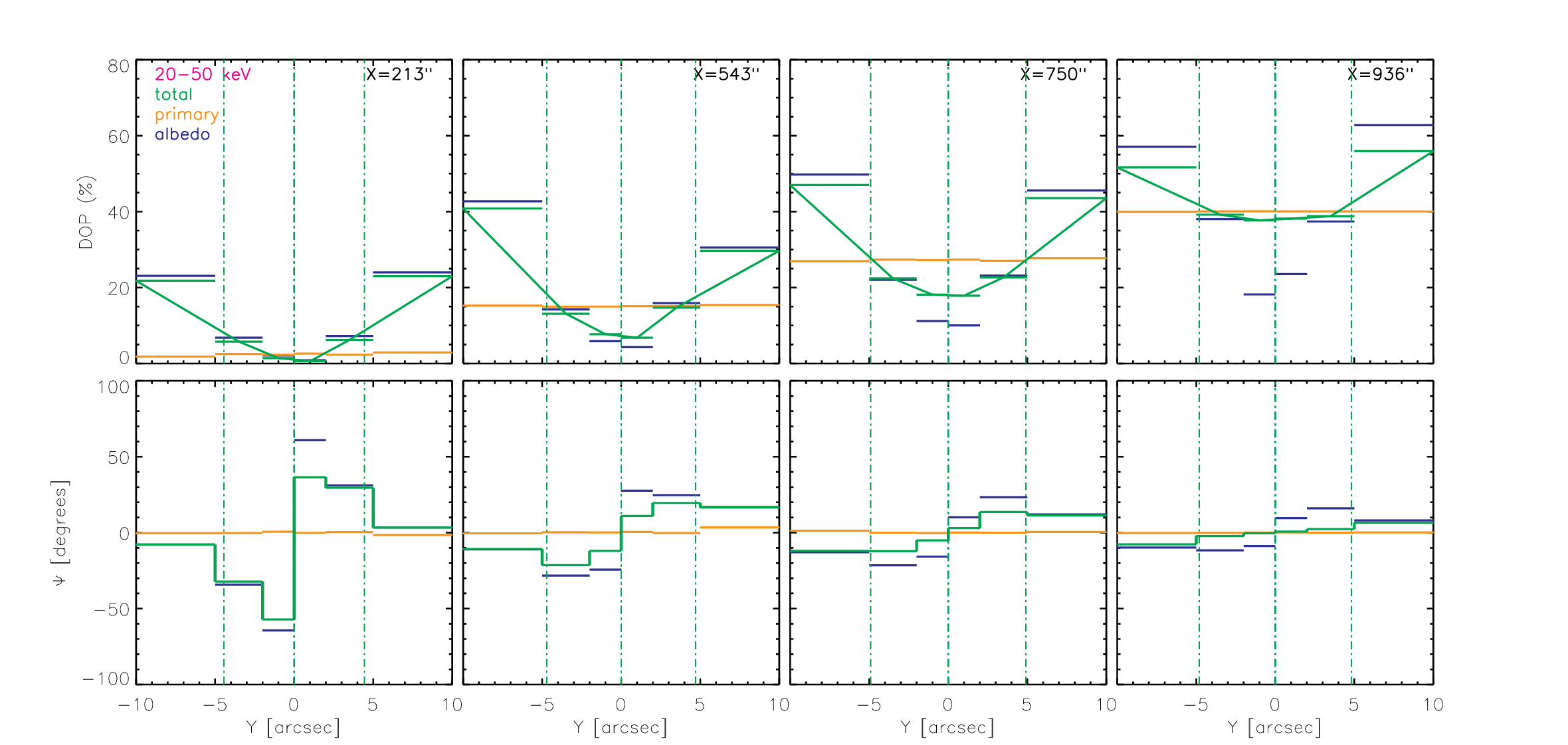}
\caption{Perpendicular to radial slices through each of the sources shown in Figure
\ref{fig:maps_dnu01} for the $\Delta\nu=0.1$ distribution.
Each of the perpendicular to radial slices are taken along $Y$  at $X=213.0'', 542.5'', 749.9'', 936.4''$ for the DOP and polarization angle $\Psi$. The lines and
colours are as in Figures \ref{fig:maps_dnu01}, \ref{fig:slicesX_dnu01}.}
\label{fig:slicesY_dnu01}
\end{figure*}

All plots for the photon distribution created by the $\Delta\nu=0.5$ electron distribution are shown in Figures (\ref{fig:maps_dnu05}-\ref{fig:slicesY_dnu05}), while Figures (\ref{fig:maps_dnu01}-\ref{fig:slicesY_dnu01}) show all plots for the photon distribution created by the $\Delta\nu=0.1$ electron distribution. Comparison of Figures \ref{fig:maps_dnu4}, \ref{fig:maps_dnu05} and \ref{fig:maps_dnu01} demonstrates that increased beaming towards the photosphere produces smaller, more concentrated and intense albedo patches.

For the $\Delta\nu=0.5$ distribution, Figure \ref{fig:slicesX_dnu05} plots the radial intensity, I (top) (along $X$ at $Y=0''$). The first two disk locations are albedo dominated, the third disk location has approximately equal contributions from the primary and albedo components and only the disk location closest to the limb is primary dominated. The primary DOP can rise as high as $\sim20\%$ at the limb and the primary $\Psi$ is radial at all locations.
Figure \ref{fig:slicesX_dnu01} (top) plots radial intensity slices (along $X$ at $Y=0''$) for the $\Delta\nu=0.1$ distribution. As expected, the first three disk locations are albedo dominated and the primary DOP can reach $\sim40\%$ at the limb. Again, the primary $\Psi$ is radial at all disk locations.

As with the quasi-isotropic $\Delta\nu=4.0$ distribution, common trends can be observed across individual sources at particular disk locations and between disk locations for both the $\Delta\nu=0.5$ and $\Delta\nu=0.1$ distributions. More importantly for observations and anisotropy deduction purposes, trends between each of the three simulated distributions ($\Delta\nu=4.0$, $\Delta\nu=0.5$ and $\Delta\nu=0.1$) can be observed, along the radial and perpendicular to radial directions, at any chosen disk location. The most notable trends are observed in the radial (X) direction, and it is these trends that may help deduce something about the anisotropy of the photon distribution for a HXR source sitting at a given disk location. Trends can be observed at all disk locations, but in this example the patterns are most noticeable in the second and third disk locations plotted. In both of these locations, the disk-centre-side DOP falls with increased beaming while the limb-side DOP rises with increased beaming.

Comparing the third disk location (for example) in Figures \ref{fig:slicesX_dnu4}, \ref{fig:slicesX_dnu05} and \ref{fig:slicesX_dnu01} shows how the radial DOP at the limb-side of the source rises with increased beaming from $\sim2\%$ for the $\Delta\nu=4.0$ distribution to $\sim18\%$ for the $\Delta\nu=0.5$ distribution to $\sim30\%$ for the $\Delta\nu=0.1$ distribution. The radial DOP at the disk-centre-side of the source falls with increased beaming from $\sim18\%$ ($\Delta\nu=4.0$) to $\sim14\%$ ($\Delta\nu=0.5$)  to $\sim4\%$ ($\Delta\nu=0.1$). The polarization angle $\Psi$ also produces similar patterns with changing anisotropy. A clear example of this can be observed by comparing the second disk location plotted in Figures \ref{fig:slicesX_dnu4}, \ref{fig:slicesX_dnu05} and \ref{fig:slicesX_dnu01}. Along the radial direction, disk-centre-side $\Psi$ generally stays at $\Psi=90^{\circ}$ for all photon anisotropies, while the outer limb-side $\Psi$ falls significantly with increased beaming, from $\Psi=60^{\circ}$ ($\Delta\nu=4.0$) to $\Psi=20^{\circ}$ ($\Delta\nu=0.5$) to $\Psi=0^{\circ}$ ($\Delta\nu=0.1$).

Therefore, the DOP and $\Psi$ patterns are a clear indication of how spatially resolved polarization measurements could be used to determine the beaming of the photon distribution. It should be noted that a (near) disk-centre source produces a slightly different trend in radial DOP with increasing photon anisotropy. The DOP at the disk-centre-side remains approximately the same for all photon anisotropies while the DOP at the limb-side falls with increased beaming (this is the opposite trend to other disk locations).

Comparing each disk location along the perpendicular to radial direction (Y) in Figures \ref{fig:slicesY_dnu4}, \ref{fig:slicesY_dnu05} and \ref{fig:slicesY_dnu01} shows that greater beaming increases the DOP over the whole extent of the source at any given location (except at the disk centre where the spatially resolved polarization along Y is approximately the same for all three distributions). This spatial increase is most noticeable at limb locations where from the source
centre to the source edge, DOP increases from $\sim0\%$ to $\sim20\%$ ($\Delta\nu=4.0$), from
$\sim20\%$ to $\sim40\%$ ($\Delta\nu=0.5$) and from $\sim40\%$ to $\sim55\%$ ($\Delta\nu=0.1$).

\subsection{High energy cutoff in the electron distribution and HXR polarization}

Spatially integrated polarization is dependent on the highest energy in the electron distribution (the high cutoff energy)
 \citep{Heristchi1987}. When calculating spatially integrated polarization,
equation (\ref{eq:DOP}) reduces to $DOP=Q$ and equation (\ref{eq:Psi}) reduces to $\Psi=0.5\arctan\left(\frac{-0}{-Q}\right)$
as $U$ sums to zero for a single measurement across the entire source. A negative DOP indicates that the polarization angle
is parallel to the radial direction ($\Psi=0^{\circ}$), while a positive DOP indicates that the polarization angle is
perpendicular to the radial direction ($\Psi=90^{\circ}$).

\begin{figure*}
\includegraphics[width=17cm]{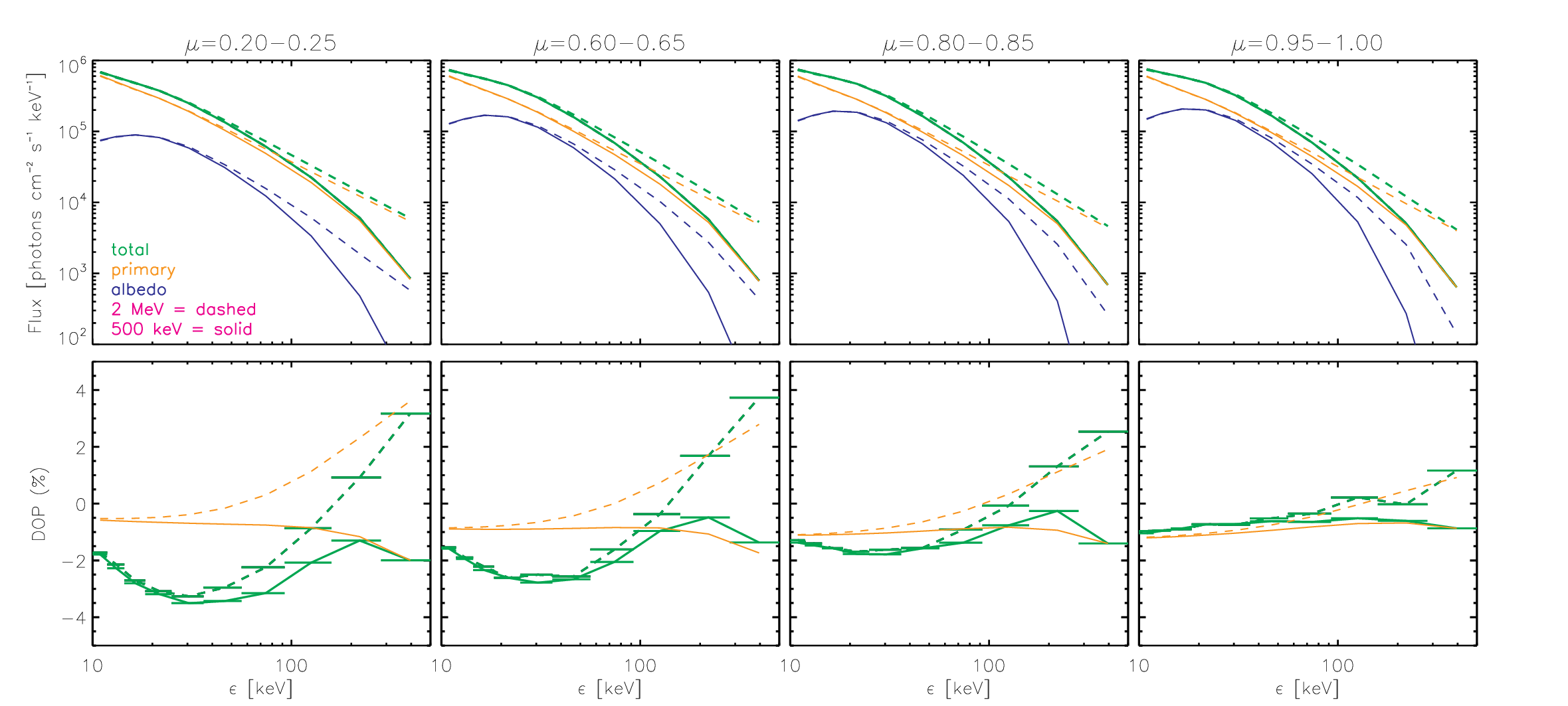}
\caption{Changes in flux (top row) and spatially integrated polarization (bottom row) for the total source (green) and primary (orange)
and albedo (blue) components with photon energy $\epsilon$ for the $\Delta\nu=4.0$ created photon distribution
with $E_{cutoff} = 500$ keV (solid lines) and $E_{cutoff} = 2$ MeV (dashed lines). Note that the
four disk locations shown are the same as those used in Figure \ref{fig:maps_dnu4}. Though there
are small changes in the magnitude (DOP) of the polarization, at high photon energies ($\sim200$
keV), the polarization direction of the $2$ MeV distribution changes from $\Psi=0^{\circ}$
(negative) to $\Psi=90^{\circ}$ (positive), while $\Psi$ for the $500$ keV distribution always
stays negative.}
\label{fig:energy_dnu4}
\end{figure*}

\begin{figure*}
\includegraphics[width=17cm]{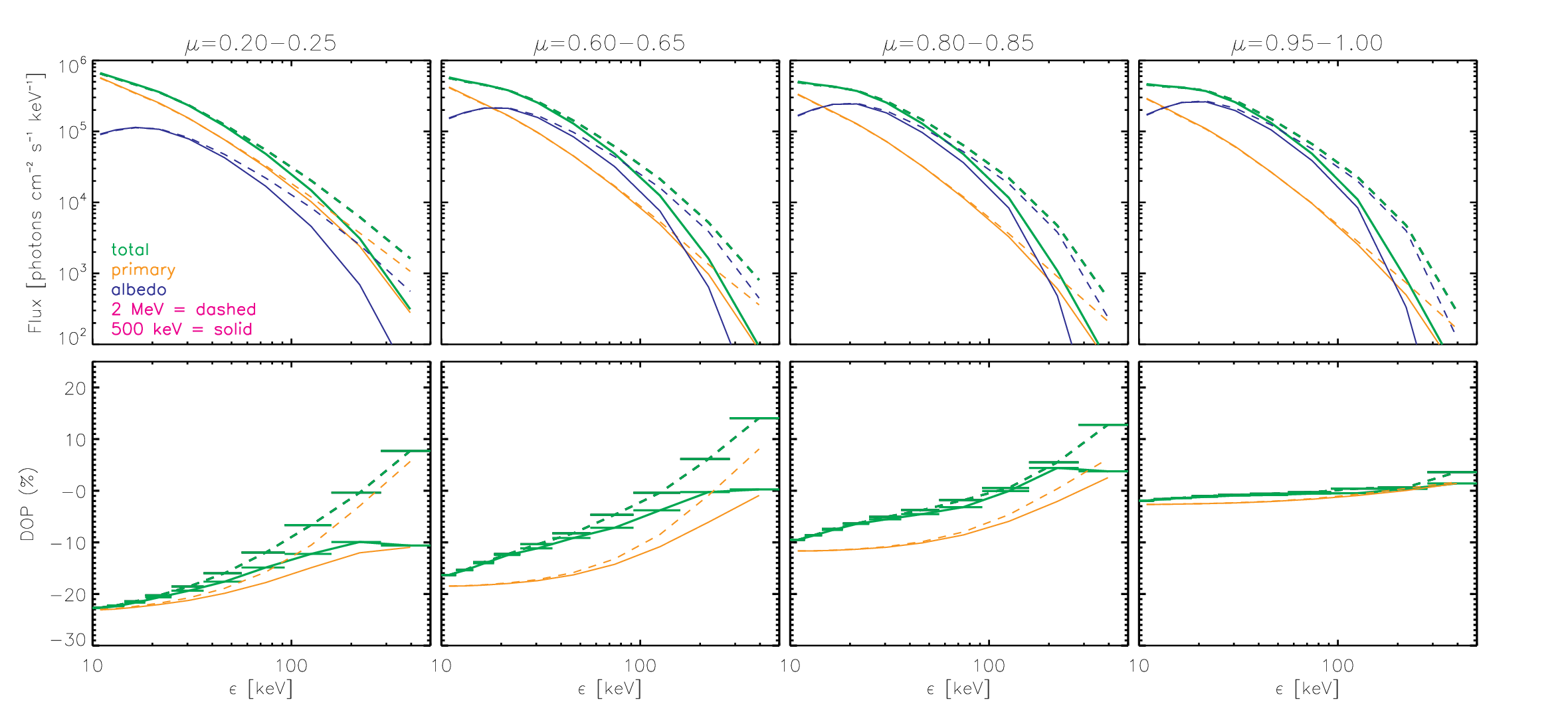}
\caption{Changes in flux (top row) and spatially integrated polarization (bottom row) for the total source (green) and primary (orange)
and albedo (blue) components with photon energy $\epsilon$ for the $\Delta\nu=0.5$ created photon distribution
with $E_{cutoff} = 500$ keV (solid lines) and $E_{cutoff} = 2$ MeV (dashed lines).}
\label{fig:energy_dnu05}
\end{figure*}

\begin{figure*}
\includegraphics[width=17cm]{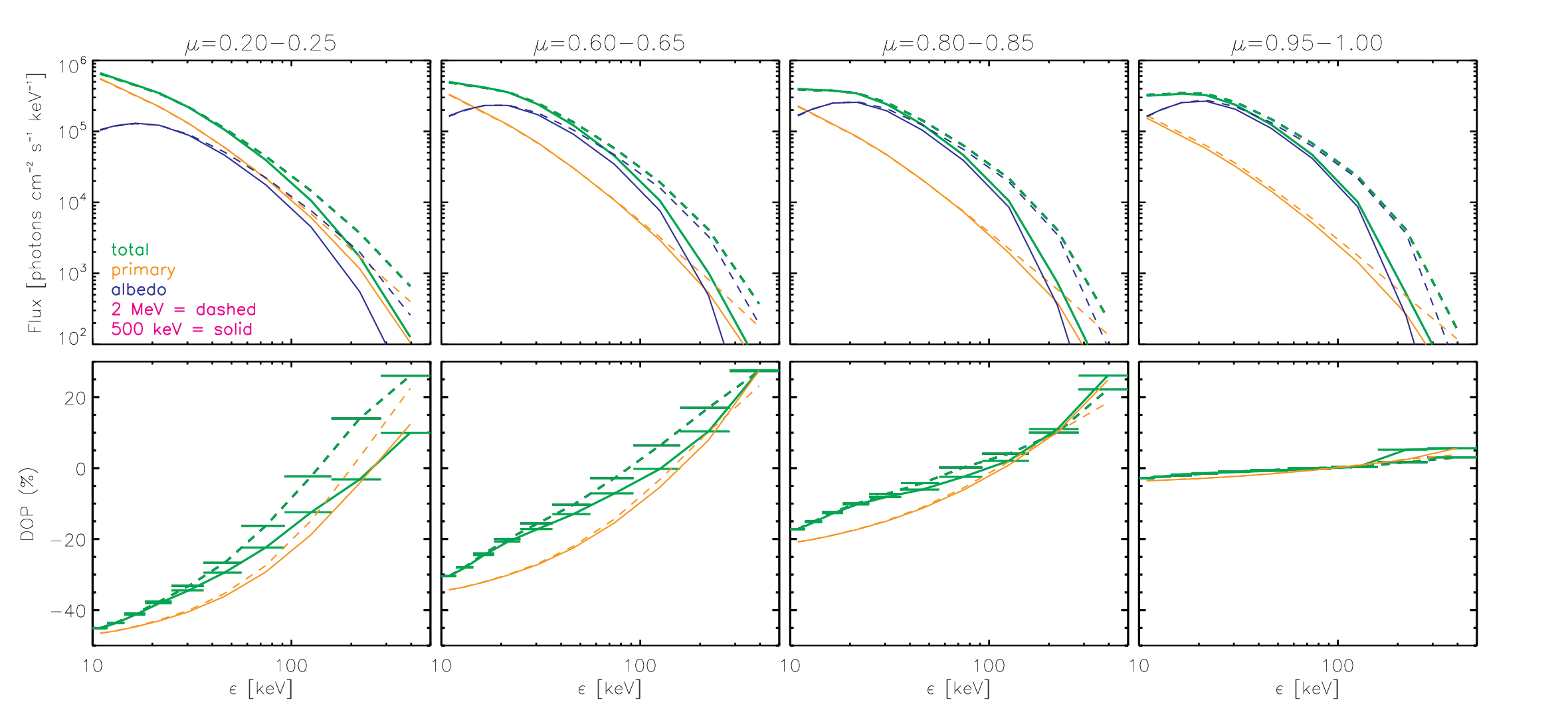}
\caption{Changes in flux (top row) and spatially integrated polarization (bottom row) for the total source (green) and primary (orange)
and albedo (blue) components with photon energy $\epsilon$ for the $\Delta\nu=0.1$ created photon distribution
with $E_{cutoff} = 500$ keV (solid lines) and $E_{cutoff} = 2$ MeV (dashed lines).}
\label{fig:energy_dnu01}
\end{figure*}

For the three electron distributions of $\Delta\nu=4.0$, $\Delta\nu=0.5$ and
$\Delta\nu=0.1$, simulations were run with two high cutoff electron energies of $E_{cutoff}=500$
keV and $E_{cutoff}=2$ MeV. Figures \ref{fig:energy_dnu4}, \ref{fig:energy_dnu05}
and \ref{fig:energy_dnu01} plot the spatially integrated polarization across the total source (green)
and the primary source only (orange) against photon energy $\epsilon$ at four disk locations
$\mu\in[0.20-0.25],[0.60-0.65],[0.80-0.85],[0.95-1.00]$ for each of the above distributions respectively. The important property to observe here is not the
magnitude but the sign of the DOP (or whether $\Psi=0^{\circ}$ or $\Psi=90^{\circ}$).

Using an electron distribution with $\Delta\nu=4.0$ (Figure \ref{fig:energy_dnu4}) and a cutoff energy of $E_{cutoff}=500$ keV
produces a photon distribution with a negative DOP at all photon energies and disk locations, while using the same distribution
with a cutoff energy of $E_{cutoff}=2$ MeV creates a photon distribution where the DOP changes from negative to positive at
$\sim100-200$ keV at all disk locations.  During bremsstahlung, in order to conserve energy, electrons with higher energies will scatter through larger angles.
When a photon is scattered through a large angle, its polarization is more likely to be directed perpendicular to the plane of emission
($\Psi=90^{\circ}$) rather than parallel to the plane of emission ($\Psi=0^{\circ}$). Therefore, a change in the direction of polarization
(from $\Psi=0^{\circ}$ to $\Psi=90^{\circ}$) indicates the presence of higher energies in the electron distribution, greater than $\sim1$ MeV.

As the beaming of the electron distribution increases (the $\Delta\nu=0.5$ and $\Delta\nu=0.1$ distributions), we see that the above statement
does not hold and it becomes more likely that both electron distributions with cutoff energies of either $E_{cutoff}=500$ keV and $E_{cutoff}=2$ MeV
will produce photons with $\Psi=90^{\circ}$. For the $\Delta\nu=0.5$ distribution (Figure
\ref{fig:energy_dnu05}), we see that as the source moves towards the solar centre, the
photon distribution created by the $E_{cutoff}=500$ keV also produces photons at higher energies (again $\sim100-200$ keV)
with $\Psi=90^{\circ}$. For the very beamed $\Delta\nu=0.1$ distribution (Figure \ref{fig:energy_dnu01}), both
the $E_{cutoff}=500$ keV and $E_{cutoff}=2$ MeV distributions produce high ($\sim100-200$ keV) energy photons with $\Psi=90^{\circ}$
at all disk locations.

Therefore using the direction of $\Psi$ as an indicator for high energies in the electron
distribution becomes less and less useful as the beaming of the photon (electron) distribution
increases. The increased beaming causes lower and lower energy photons to scatter at larger angles, especially at locations closer to the solar centre,
hence producing photons with $\Psi=90^{\circ}$ in the $E_{cutoff}=500$ keV distributions.
Note that this method is only useful when the beaming of the photon distribution has already been determined.

\section{Discussion and conclusions}
The simulation results show that Compton backscattering produces a clear and distinct albedo polarization pattern across an HXR source at energies of 20-50 keV. Trends can be observed for both of the measured polarization parameters, DOP and $\Psi$, and are clear indications of the existence of an albedo component in comparison with the constant, radial polarization of the primary emission. This means that spatially resolved polarization can be used to probe structure within HXR footpoint sources, helping us distinguish between the bremsstrahlung source and the albedo source. More importantly, at a single disk location, spatially resolved DOP and $\Psi$ exhibit clear variations with changing photon directivity, along both the radial and perpendicular to radial directions and can be used to determine the anisotropy of the electron distribution.
Therefore, to take advantage of these properties requires reliable polarization measurements with an angular resolution of $\sim5''-10''$ over the peak albedo energies of $\sim20-50$ keV.

The simulations also suggest that for $\sim $ isotropic HXR sources, spatially integrated polarization angle measurements, $\Psi$, from low to high energies,
with consideration of the changes due to albedo, could help indirectly infer the highest energy in
the electron distribution. For near isotropic sources implied by RHESSI X-ray observations
\citep{Kasparovaetal2007,KontarBrown2006}, changes in $\Psi$ from the radial ($\Psi=0^{\circ}$) to the perpendicular to radial direction ($\Psi=90^{\circ}$) may help indicate the presence of high energy electrons ($\sim\ge1$ MeV) present in the electron distribution, if the
photon anisotropy is known. Changes in spatially integrated DOP measurements, from low to high energies, will also help determine the anisotropy of the photon distribution.

Currently, when observing solar flares we do not have the instrumentation required to produce spatially resolved polarization measurements. It is doubtful whether {\it near} future missions such as the Gamma-Ray Imager/Polarimeter for Solar Flares (GRIPS) \citep{Shihetal2009} may have the capability to measure polarization over albedo energies, even though it should be able to measure polarization across $12.5''$. The best polarization measurements are likely to be over the range of 150-650 keV with a $\sim$4$\%$ minimum detectable polarization \citep{Shihetal2009}. At energies greater than $\sim$100 keV, the albedo flux drops off steeply, so it is very unlikely that any albedo component could be detected at these energies and the polarization across the observed HXR source would only be from the bremsstrahlung emission. Therefore, good polarization measurements at 150-650 keV from flares with high fluxes will give us a direct measurement of the primary component and DOP/$\Psi$ measurements at these energies may be used to infer the high energy cutoff in the electron distribution, for relatively isotropic distributions.

\begin{acknowledgements}
NLSJ and EPK are grateful to A. Gordon Emslie for the use of his bremsstrahlung cross-section code
and Hugh Hudson, Albert Shih and Gordon Holman for useful comments and discussions. NLSJ is funded by STFC and SUPA
scholarships. EPK gratefully acknowledges the financial support by the Leverhulme Trust (EPK), STFC
Rolling Grant, and by the European Commission through the HESPE (FP7-SPACE-2010-263086) Network
acknowledged.
\end{acknowledgements}

\appendix
\section{Compton scattering and new scattering angles and energies}\label{A.1.2}

In the Monte Carlo simulations, when a Compton scattering occurs, the properties of the outgoing photon: energy $\epsilon$, polar
scattering angle $\theta_{S}$, azimuthal scattering angle $\phi_{S}$ and the two linear
polarisation parameters $Q$ and $U$ need to be updated. New polar scattering angles $\theta_{S}$
for each photon can be found by integrating the Klein-Nishina differential cross section over
$\phi_{S}$ to produce a differential cross-section that is only dependent on $\epsilon$ and
$\theta_{S}$. This is the Klein-Nishina \citep{KleinNishina1929} differential cross-section for a
completely unpolarised isotropic beam and is given by,
\begin{equation}
\frac{d\sigma_{c}(\theta_{S},\epsilon)}{d\Omega}=\frac{1}{2}r_{0}^{2}\left(\left(\frac{\epsilon}{\epsilon_{0}}\right)^{3}+
\frac{\epsilon_{0}}{\epsilon}-\left(\frac{\epsilon}{\epsilon_{0}}\right)^{2}\sin^{2}\theta_{S}\right).
\end{equation}
Since MC simulations operate by drawing numbers randomly from a given distribution, this means
that $\theta_{S}$ can be easily found by matching each value of
$\theta_{S}\in[0^{\circ},180^{\circ}]$ with a number $\zeta_{\theta}\in[0,1]$ for every value of
$\epsilon$ using,
\begin{equation}
\zeta_{\theta}=\frac{2\pi}{\sigma_{c}}\int_{0}^{\theta_{S}}\frac{d\sigma_{c}(\theta_{S},\epsilon)}{d\Omega}d\theta_{S},
\end{equation}
where $\sigma_{c}$ is the total Compton scattering cross-section. New values of $\theta_{S}$ are simply
drawn at each scattering using the photon energy before scattering and a random, uniform number between
$0$ and $1$.

Once the new scattering angles $\theta_{S}$ are obtained then the new photon energy $\epsilon$ can
be easily found using,
\begin{equation}
\epsilon=\frac{\epsilon_{0}}{1+\frac{\epsilon_{0}}{mc^{2}}(1-\cos\theta_{S})}.
\end{equation}

\section{Compton scattering and photon polarization states} \label{app_B}
If the photon distribution is completely isotropic and unpolarised then the azimuthal scattering
angle $\phi_{S}$ can just be sampled from a uniform distribution between $0$ and $2\pi$, but this
is not true for the more general polarization dependent case. The probability density function of
obtaining a value of $\phi_{S}$ between $\phi_{S}$ and $\phi_{S}+d\phi_{S}$ can be described by:
\begin{equation*}
P(\phi_{S})=\frac{1}{2\pi}\frac{d\sigma_{c}(\epsilon,\theta_{S},\phi_{S})/d\Omega}{d\sigma_{c}(\epsilon,\theta_{S})/d\Omega}=
\end{equation*}
\begin{equation}
=\frac{1}{2\pi}\frac{\frac{\epsilon_{0}}{\epsilon}+\frac{\epsilon}{\epsilon_{0}}-
\sin^{2}\theta_{S}(1-Q\cos2\phi_{S}-U\sin2\phi_{S})}{\frac{\epsilon_{0}}{\epsilon}+\frac{\epsilon}{\epsilon_{0}}-\sin^{2}\theta_{S}}
\end{equation}
with the maximum value of this function given by:
\begin{equation}
P_{max}(\phi_S)=\frac{1}{2\pi}\frac{\frac{\epsilon_{0}}{\epsilon}+\frac{\epsilon}{\epsilon_{0}}-
\sin^{2}\theta_{S}\left(1-\sqrt{Q^{2}+U^{2}}\right)}{(\frac{\epsilon_{0}}{\epsilon}+
\frac{\epsilon}{\epsilon_{0}}-\sin^{2}\theta_{S})}
\end{equation}
Firstly, a value of $\phi_{S}$ is sampled between $0$ and $2\pi$. The condition that
$P(\phi_S)<P_{max}(\phi_S)$ is then used to accept a value of $\phi_{S}$ and provides a method for
sampling values of $\phi_{S}$ for each photon, using the new values $\theta_S$ and $\epsilon$
already calculated for each photon/photon packet. This method is repeated until the condition is
satisfied for each photon and each photon is provided with an azimuthal scattering angle
\citep{Salvatetal2008}.

Due to Compton scattering, the Stokes parameters have to be updated using the scattering matrix T
\citep{McMaster1961, BaiRamaty1978}				\begin{equation} 		T(\epsilon,\theta_{S})= 		
\left( 		\begin{array}{ccc}
  		\frac{\epsilon}{\epsilon_{0}}+\frac{\epsilon_{0}}{\epsilon}+\sin^{2}\theta_{s}&
  \sin^{2}\theta_{s}  &0   \\
  		\sin^{2}\theta_{s}&\cos^{2}\theta_{s}+1   & 0  \\
  		0& 0  &2\cos\theta_{s}
		\end{array} 		\right)
\label{eq:T_matrix}
		\end{equation} Before a scattering, the Stokes parameters have to be rotated by $\phi_S$ so
that they are defined relative to the plane of scattering. The Stokes parameters before scattering
are rotated by the rotation matrix $M(\phi_{S})$ given by:
		
		\begin{equation} 		M(\phi_{S})= 		\left( 		\begin{array}{ccc} 		1&0&0 \\
  		0 &\cos2\phi_{S}  &\sin2\phi_{S} \\
  		0 &-\sin2\phi_{S}  &\cos2\phi_{S} \\
		\end{array} 		\right) 		\end{equation} After a scattering, the Stokes
parameters have to be rotated again so that they are defined relative to the starting position of
the source. The Stokes parameters after scattering are rotated by the rotation matrix
$M(-\Gamma)$.
		
		\begin{equation} 		M(-\Gamma)= 		\left( 		\begin{array}{ccc} 		1&0&0 \\
  		0 &\cos-2\Gamma  &\sin-2\Gamma \\
  		0 &-\sin-2\Gamma  &\cos-2\Gamma \\
		\end{array} 		\right), 		\end{equation}
\begin{figure}
\includegraphics[width=75mm]{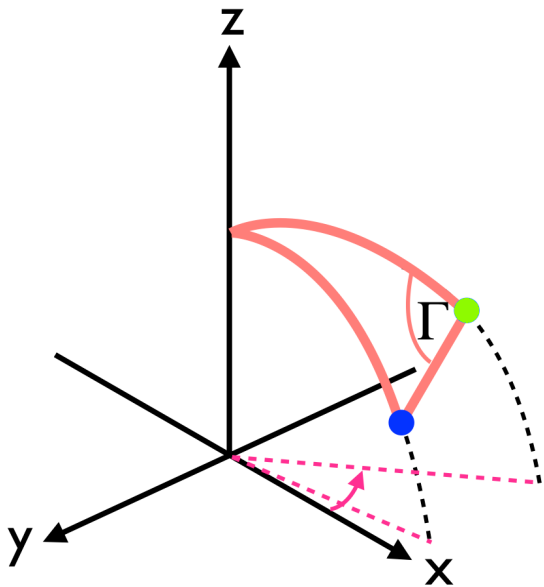}
\caption{The position of the photon before scattering (blue) and after scattering (green)
and the angle $\Gamma$ that determines the final rotation of the Stokes parameters back into the
frame of the source from the scattering frame.}
\label{fig:rotation}
\end{figure}
where
\begin{equation}
\cos\Gamma=\pm\frac{w'-\cos\theta_{s}w}{\sqrt{1-\cos^{2}\theta_{s}}\sqrt{1-w^2}}.
\label{eq:cosGamma}
\end{equation}
$w$ and $w'$ are the current and previous $z$ direction cosines respectively \citep{Hovenier1983}.
The $\pm$ in equation (\ref{eq:cosGamma}) is present due to the negative sign being used when
$\pi\leq\phi_{s}\leq2\pi$ and the positive sign for $0\leq\phi_{S}<\pi$. The angle $\Gamma$ is
shown in Figure \ref{fig:rotation}. $\Gamma$ is the angle between the scattering plane and the
normal plane in the frame of the source. Therefore during a Compton scattering the order of the
rotations on the Stokes vector $[I Q U]$ is $M(\phi_{S})T(\epsilon,\theta_{S})M(-\Gamma)$.

\bibliographystyle{aa}
\bibliography{spatpol_references}

\end{document}